\documentclass[twocolumn, 8pt]{article}
\usepackage{usenix2019_v3}

\usepackage{tikz}
\usepackage{amsmath}
\usepackage{algorithm}
\usepackage{algorithmic}
\usepackage{filecontents}
\usepackage{graphicx}
\usepackage{listings}
\usepackage{bussproofs}
\usepackage{varwidth}
\usepackage{amssymb}
\usepackage[makeroom]{cancel}
\usepackage{xspace} \usepackage{xargs}
\usepackage{todonotes}
\usepackage{soul}
\usepackage{url}
\usepackage[numbers]{natbib}

\newtheorem{definition}{Definition}[section]
\newtheorem{example}{Example}[section]

\newcommandx{\todos}[2][1=]{\todo[inline,caption={},linecolor=red,backgroundcolor=red!25,bordercolor=red,#1]{\textbf{TODO:
}#2}}
\newcommandx{\commentJames}[2][1=]{\todo[inline,caption={},linecolor=green,backgroundcolor=green!25,bordercolor=green,#1]{\small#2\textbf{\\--
James}}\xspace}
\newcommandx{\commentJamesM}[2][1=]{\todo[caption={},linecolor=green,backgroundcolor=green!25,bordercolor=green,#1]{\small#2\textbf{\\--
James}}\xspace}
\newcommandx{\commentAlberto}[2][1=]{\todo[inline,caption={},linecolor=cyan,backgroundcolor=cyan!25,bordercolor=cyan,#1]{\small#2\textbf{\\--
Alberto}}\xspace}
\newcommandx{\commentAlbertoM}[2][1=]{\todo[caption={},linecolor=cyan,backgroundcolor=cyan!25,bordercolor=cyan,#1]{\small#2\textbf{\\--
Alberto}}\xspace}

\begin{document}
%-------------------------------------------------------------------------------

%don't want date printed
\date{}

\title{\Large \bf Bitcoin Trace-Net:\\
Formal Contract Verification at Signing Time}

%for single author (just remove % characters)
\author{
{\rm \Large{James Chiang}}\\
\\
{\rm \normalsize{Technical University of Denmark}} \\
{\rm \normalsize{Kongens Lyngby, Denmark}} \\
{\rm \normalsize{\texttt{jchi@dtu.dk}}}
} % end author

\maketitle

%-------------------------------------------------------------------------------
\begin{abstract}
%-------------------------------------------------------------------------------
Smart contracting protocols promise to regulate the transfer of cryptocurrency amongst participants in a trustless manner. A safe smart contract implementation should ensure that each participant can always append a contract transaction to the blockchain in order move the contract towards secure completion. To this goal, we propose Bitcoin Trace-Net, a contract verification framework which generates an executable symbolic model from the underlying contract implementation. A Trace-Net model consists of a Petri Net formalism enriched with a Dolev-Yao-like actor knowledge model. The explicit symbolic actor knowledge model supports the verification of contracts featuring cryptographic sub-protocols, which may not be observable on the blockchain. Trace-Net is sufficiently expressive to accurately model blockchain semantics such as the delay between a transaction broadcast and its subsequent confirmation, as well as adversarial blockchain reorganizations of finite depths, both of which can break smart contract safety. As an implementation level framework, Trace-Net can be instantiated at run-time to monitor and verify smart contract protocol executions.
\end{abstract}
%-------------------------------------------------------------------------------
\section{Introduction}
%-------------------------------------------------------------------------------
The term "Smart contracts" is commonly attributed to Nick Szabo \cite{szabo1997idea}, who is known for his early work on digital cash and digital contracts. Smart contracts promise contract implementations between actors to be executed in a trustless manner without reliance on any trusted intermediaries. To this goal, blockchain protocols offer both permissionless execution and immutability of transactions. These properties promise the realization of trustless smart contracts which are safely executable by any contract participant and robust against any adversarial actions by counter-parties. More formally, we are interested in verifying contracts for execution traces which lead to safe outcomes. If the state space of the contract is finite, such a safety notion expressed as a trace property is decidable. In practice, many smart contracts are implemented in general-purpose Turing-complete languages and thus cannot be definitively verified for trustless execution. Instead, substantial attention has been directed towards the analysis of vulnerability patterns \cite{atzei2016survey} on more expressive contracting platforms such as Ethereum. Static analysis of smart contracts to avoid vulnerability patterns \cite{grishchenko2018foundations} can be highly effective in finding bugs. However they cannot prove the absence of any unintentional contract behaviour and therefore cannot fully deliver on the original promise of trustless smart contract execution.

Contracts implemented in Bitcoin are famously Turing-incomplete and feature finite state-spaces due to its limited expressiveness of the underlying Script language. Bitcoin as a contracting platform therefore lends itself to verification with model-checking approaches. This has been thoroughly investigated by Bartoletti \& Zunino with BitML \cite{bartoletti2018bitml} \cite{atzei2019developing} \cite{bartoletti2020renegotiation} \cite{bartoletti2019liquidity} \cite{bartoletti2020bitcoin}, a higher-level contract specification language for Bitcoin which compiles to Bitcoin transactions, appendable to the Bitcoin blockchain by the contracting participants. BitML can be symbolically executed at the language level and provides safe compiler guarantees so that honest actors can always move the specified contract forward during execution: This is a big step towards secure smart contract development since safety properties can be guaranteed at design time. However, contract specification languages at a higher abstraction level come at the cost of expressiveness when compared to designing contracts at the raw Bitcoin transaction level: Implementing smart contracts directly at the computational level is unwieldy, can also enable optimizations for higher on-chain privacy or lower execution costs. For contracting protocols such as atomic swaps \cite{herlihy2018atomic}, lightning \cite{poon16bitcoin} or state-channels \cite{aumayr2020generalized}, where on-chain transaction privacy or cost efficiency are paramount, BitML may not be the appropriate implementation and verification tool.

Instead, Trace-Net is intended as an implementation level verification framework. As such, we also emphasize the need to accurately model blockchain semantics such as the delay between a transaction broadcast and its confirmation as well as possible blockchain reorganizations \cite{judmayer2019pay} \cite{moroz2020double} executed by the adversary, both of which can break the security of a smart contract design. Recent protocol specification errors in the Bitcoin Lightning \cite{lightningcve} protocol, for example, also indicate that the verification of contracts developed at the Bitcoin implementation level remain an open challenge. Bitcoin contracting protocols have been verified with cryptographic frameworks \cite{canetti2001uc}, but manual security proofs cannot be automated for different contract designs.

\paragraph{Contributions} We introduce Trace-Net as an automated contract verification framework applicable at the the transaction level of UTXO-based blockchain protocols. Trace-Net generates an executable contract model from the symbolic execution of unsigned transactions implementing an interactive contract protocol: This symbolic contract model consists of an extended Petri Net formalism representing on-chain contract states and an extensible Dolev-Yao-like \cite{dolev1983security} actor knowledge model, which expresses the actor's ability to derive knowledge from observations and publicly known cryptographic functions. Modeling actor knowledge explicitly allows Trace-Net to be extended with additional cryptographic sub-protocols which may be unobservable on-chain, such as adaptor signatures \cite{aumayr2020generalized} and can therefore support a wide range of contract protocol designs. Furthermore, separate modeling of on-chain and actor knowledge states enable Trace-Net to express blockchain reorganizations which may occur during contract execution: Such an event will reorder transactions whilst retaining the states of each actor's knowledge. Blockchain reoganizations can frequently occur on blockchains with less security and contracts may need to be verified against such events. To the best of our knowledge, Trace-Net is the first automated approach enabling the verification of safety properties in the presence of adversarial blockchain reorganizations.

The Trace-Net contract model is unfolded into a finite-state transition system, for which temporal properties are decidable: In particular we introduce a formalisation of the \emph{trustless execution} property, which intuitively holds true if a contract can be safely terminated by the verifying agent despite any adversarial strategies executed by the counter-party. The positive verification of the trustless execution property results in the identification of all safe strategies for the verifying actor, which can be enforced at run-time. The verification of this property can be extended to determine the safety of contract updates which may occur during contract execution. Furthermore, we introduce the formal notion of \emph{contract stability}, which describes the safety of remaining in an intermediary, non-terminal contract state: This is a useful notion for the verification of off-chain contracts such as state-channels, where the termination of a live contract protocol is deferred so that it can be updated at a later point in time.

We believe Trace-Net introduces the following promising implications: Firstly, given a higher level specification of the contract, Trace-Net can be instantiated to ensure implementation correctness, by verifying that trace properties of the specification are also present at the implementation level. Of course, a contract protocol can also be verified against general safety policies expressed as temporal, trace properties, such as \emph{trustless execution} and \emph{contract stability}. Thirdly, the ability to perform automated verification at run-time can be useful for monitoring contract protocol executions. Dedicated key management systems in secure execution environments tasked with the secure signing of transactions could enforce universal safety policies at signing-time: Otherwise, safe signing approaches must rely on contract-specific transaction pattern matching \cite{devrandom2019} which cannot be generalized to other contract types and cannot formally guarantee any contract safety properties. 

\paragraph{Organization} The rest of the paper is organized as follows. In section \ref{Related Work} we summarize existing work relevant to our solution but also provide a wider overview of approaches to formal correctness in smart contract design. In section \ref{A Symbolic Model of Bitcoin Transactions} we introduce our model of Bitcoin transactions and their symbolic execution. Subsequently, the Trace-Net framework (section \ref{Trace-Net: A Symbolic Model of Contracts}) is introduced which consists of actor knowledge models (\ref{Actor Knowledge}, \ref{Cryptographic Extensions}) and an on-chain Petri Net \ref{Petri Net Output State Model} formalism derived from the previous symbolic execution of the underlying transactions. Finally, sections \ref{section: Automated Trace-Net Analysis} and \ref{Contract Safety Properties} are dedicated to the automated generation and analysis of the Trace-Net model for security properties such as \emph{trustless execution}, \emph{contract update safety} and \emph{contract state stability}.

%-------------------------------------------------------------------------------
\section{Related Work} \label{Related Work}
%-------------------------------------------------------------------------------

Efforts to provide correctness guarantees in smart contracts inherently face trace-offs between the expressiveness of the implementation language and the decidability of the security properties of interest. In this section, we aim to provide a brief overview of the spectrum of formal frameworks and languages intended to ensure contract correctness and safety.

\paragraph{Analysis of Bitcoin Script} The Bitcoin Script language is non-Turing-complete and often used as an implementation language for various contracting protocols such as atomic swaps \cite{herlihy2018atomic}, coin join \cite{maxwell2013coinjoin} variants, payment channel networks \cite{poon16bitcoin}, generalized state channels \cite{aumayr2020generalized} and other contract designs surveyed by past literature \cite{bartoletti2017empirical}. However, even though the limited expressiveness of Bitcoin Script is often claimed to facilitate the manual investigation of its behaviour, only a fragment of its language has been formalized (Klomp \& Bracciali \cite{klomp2018symbolic}), which makes symbolic execution and formal verification of Bitcoin Script challenging. Miniscript \cite{wuille2019miniscript} \cite{wuille2019cppminiscript} \cite{poelstra2019rustminiscript} is a practical framework that enables symbolic execution of composable Bitcoin Script fragments, and is used by Trace-Net to lift a symbolic model from raw contract transactions. We note, however, that it is ultimately difficult to reason about the correctness of Miniscript given that the underlying Bitcoin Script semantics are only informally described. Nonetheless, developing contracts at the Bitcoin Script and transaction level provides the flexibility of optimizing the on-chain footprint for privacy and reduced execution cost.

\paragraph{Contracts as Cryptographic Protocols} Analysis of individual Bitcoin output scripts remains insufficient to reason about the execution of interactive Bitcoin contracts implemented over \emph{multiple} transactions. The cryptography community has successfully instantiated the Universal Composability \cite{canetti2001uc} framework to manually prove the security of interactive contracts designs \cite{malavolta2019anonymous} \cite{aumayr2020generalized}. Alternatively, Andrychowicz et al. have proposed to model contract participants as timed automata \cite{andrychowicz2014timedautomata} executing an interactive protocol, such that possible contract executions can be exhaustively explored in a model checker such as UPPAAL. This latter approach, however, still requires a contract model to manually specified. Trace-Net builds upon past work on both symbolic Bitcoin Script execution and the treatment of Bitcoin contracts as interactive, cryptographic protocols and offers the ability to automatically generate a symbolic, executable model from raw transactions.

\paragraph{FSM Contract Languages} \emph{BitML} by Bartoletti and Zunino \cite{bartoletti2018bitml} \cite{atzei2019developing} \cite{bartoletti2020renegotiation} \cite{bartoletti2019liquidity} \cite{bartoletti2020bitcoin} is a high-level contracting language for Bitcoin which features safe execution guarantees provided by compiler semantics. BitML is an elegant process algebra intended for the implementation of smart contracts, but can also be executed at the symbolic language level, providing the contract designer a secure tool to ensure design intent and contract implementation are in agreement. Furthermore, executable BitML expressions are highly amenable to the verification of general safety properties, such as liquidity \cite{bartoletti2019liquidity}, which ensures that funds are never locked inside a contract design. Because the symbolic execution of BitML expressions can always be unfolded into a finite state transition system, we denote BitML as a FSM contract language, over which temporal properties are decidable. The higher level BitML abstraction and its security guarantees come at the cost of reduced control of the resulting on-chain footprint during contract execution, which may affect its usefulness in applications which emphasize on-chain privacy or execution cost. 

We note that general-purpose Turing-complete languages such as Solidity can also be represented in FSM-like form, making it amenable to verification. \emph{FSolidM} by Mavridou and Laszka \cite{mavridou2018fsm} \cite{mavridou2018tool} \cite{mavridou2019verisolid} is a high-level contract language which translates to Solidity. FSolidM is an adapted form of the Behaviour, Interaction and Priority (BIP) framework \cite{noureddine2014reduction} and aims to introduce rigorous control state and contact execution transition semantics. This has the potential of making the verification of some safety properties decidable, in particular temporal properties over user-defined control states using symbolic execution and predicate abstraction techniques.

\paragraph{Typed Functional Languages} Strong typing systems can enforce the absence of many generalized vulnerability patterns. Furthermore, functional languages lend themselves towards translation into theorem proving frameworks, where assisted proofs for contract properties can be performed. Simplicity by O'Connor \cite{oconnor2018simplicity} is a typed functional contract language intended for UTXO-blockchains such as Bitcoin, with execution semantics formalized in Coq. Scilla \cite{sergey2018scilla} \cite{sergey2019safer} draws inspiration from both functional and automata-based languages, but ultimately represents a functional language with an adaptor to Coq. Importantly, Scilla implements inter-contract communication as functions without side-effects: A contract in an account-based blockchain platform can only call an external contract at the very end of a function execution, so that no external effects are possible in the midst of a function, thereby preventing reentrancy \cite{bartoletti2017empirical} vulnerabilities and also facilitating the implementation of inter-blockchain contract messaging in sharded blockchains.

\paragraph{Detection of Vulnerability Patterns} Both static analysis \cite{luu2016oyente} \cite{tsankov2018securify} \cite{permenev2020verx} \cite{schneidewind2020ethor} and theorem proving approaches \cite{hirai2017defining} \cite{bhargavan2016formal} \cite{hildenbrandt2017kevm} \cite{grishchenko2018semantic} have been successfully deployed to detect generalized vulnerabilities previously exploited on the Ethereum blockchain. Vulnerability patterns frequently arise when there is a mismatch between the intuition of the developer and the underlying execution semantics of the blockchain. In Ethereum, common vulnerability patterns include the reentrancy \cite{atzei2016survey} problem, transaction ordering, block timestamp dependency, call stack depth limitation, unchecked send and suicidal contracts \cite{nikolic2018finding}. Static analysis tools however, cannot guarantee the absence of bugs, nor can they guarantee contract safety, although VerX \cite{permenev2020verx} has demonstrated decidability of contract-specific temporal properties for a fragment of the EVM language.

\paragraph{Domain-specific Contract Languages} Finally, we note that domain-specific contracting languages have been proposed for specific contract archetypes. The financial contract language from Bahr et al. \cite{bahr2015certified}, for example, provides very simple expressions for traditional financial contracts formalized by Peyton et al. \cite{jones2001composing}. These include foreign exchange swaps and options, credit default swaps, and portfolios holding contracts with multiple counter-parties. Marlowe \cite{seijas2018marlowe} provides a language to compile such contract types à Peyton et al. to the Cardano blockchain run-time. “Trustless” contract execution of such contracts is guaranteed by the execution framework \cite{egelund2017automated} targeting the EVM run-time, though formalized compiler semantics do not exist to the best of our knowledge. Findel \cite{biryukov2017findel} is another DSL for financial derivatives executed on Ethereum. We presume that additional domain-specific contracting languages will emerge as blockchain platforms and smart contract applications mature.

%-------------------------------------------------------------------------------
\section{Symbolic Execution of Bitcoin Transactions} \label{A Symbolic Model of Bitcoin Transactions}
%-------------------------------------------------------------------------------

% We provide a simplified model of Bitcoin transactions for our purposes of instantiating the Trace-Net framework. The initial paragraph below outlines the anatomy of a Bitcoin transaction, while subsequent paragraphs in this section will further detail the semantics of transferring funds and transaction time-locks, which enforce an earliest blockheight in the blockchain at which the transaction can be amended.

We begin this section with a simplified model of Bitcoin transactions and their general execution for the general reader. In the subsequent section \ref{Symbolic Execution of Transactions}, we introduce the symbolic execution of transaction scripts with Miniscript semantics (\ref{Miniscript}), resulting in symbolic execution paths for each transaction output. Section \ref{Transaction Confirmation Semantics} outlines accurate transaction confirmation on the blockchain, which requires finite delays between the initial transaction broadcast and its inclusion in the blockchain: We do not model transaction confirmations as final, but account for a possibility of blockchain reorganizations of finite depths by an adversarial actor.

\begin{figure}
\includegraphics[width=0.47\textwidth]{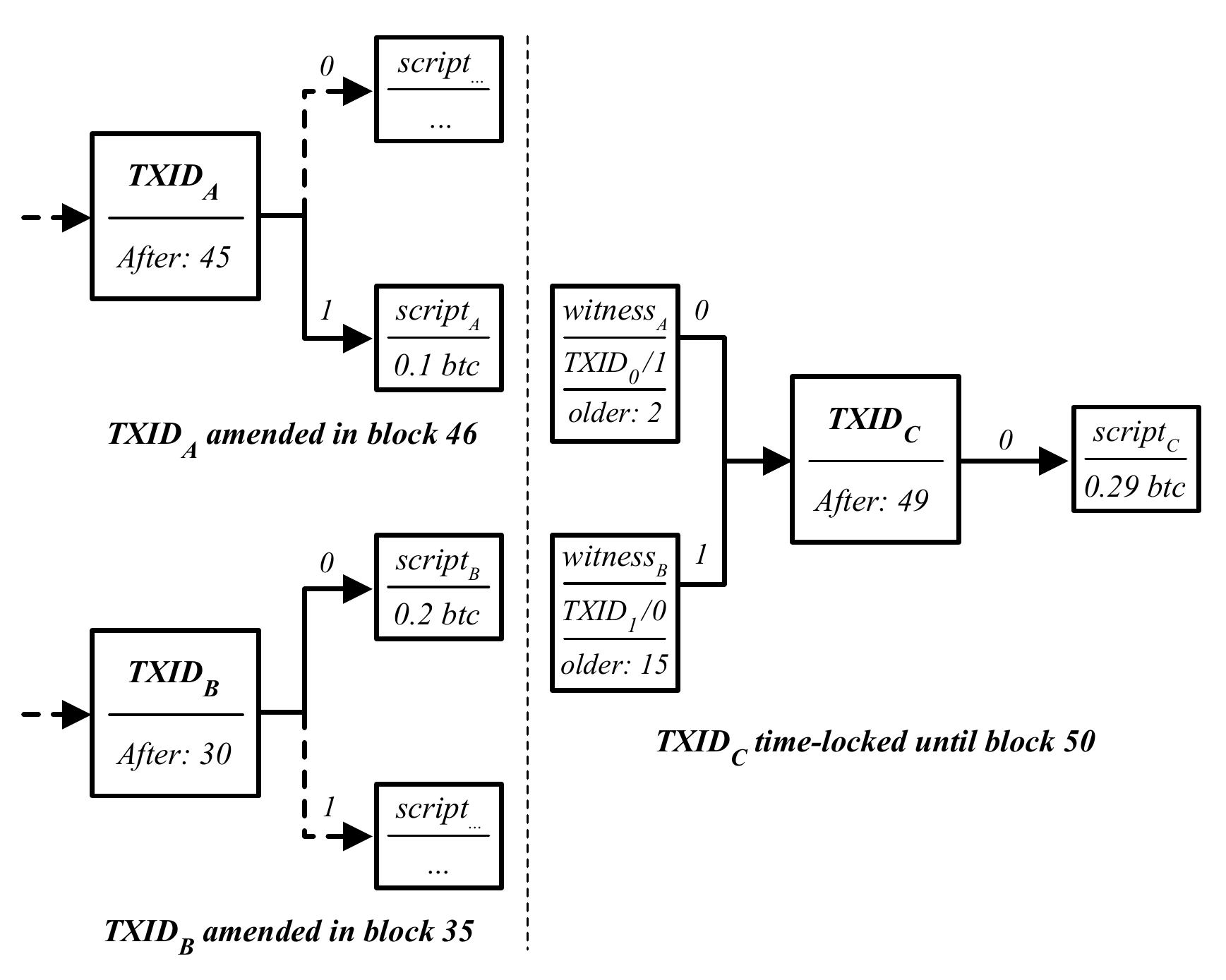} 
\caption{Transaction C (\textit{$TXID_C$}) spends outputs at indices 1 \& 0 from transactions A \& B respectively. Transaction C is time-locked until blockheight 50 since it is limited by the older time-lock of input 1: Note that the older time-lock of input 0 in transaction C releases earlier at blockheight of 48.}
\label{fig:tx_anatomy}
\end{figure}

\paragraph{Transaction Anatomy} A Bitcoin transaction consists of the attribute tuple $(tx.\textsf{id}, tx.\vec{\textsf{ins}}, tx.\vec{\textsf{outs}}, tx.\textsf{after})$, where $\vec{\textsf{ins}}$ and $\vec{\textsf{outs}}$ are input and output vectors respectively. Transaction \emph{outputs} represent funds which are spendable. A single transaction \emph{input} always refers to a single output of a previous transaction in the blockchain, thereby forwarding or \emph{spending} its value to outputs of the spending transaction (figure \ref{fig:tx_anatomy}). 

The $tx.\textsf{id}$ attibute is unique to each transaction and is obtained by hashing the transaction byte string in network serialization form which encodes the \emph{inputs}, \emph{outputs} and \emph{after} attributes. The $tx.\textsf{after}$ transaction attribute encodes a transaction level time-lock and is detailed in the subsequent \emph{transaction time-locks} paragraph.

Each input in a Bitcoin transaction is represented by an attribute tuple $(in.\textsf{prevout}, in.\textsf{witness}, in.\textsf{older})$. The $in.\textsf{prevout}$ attribute uniquely references the redeemed output in an output vector of a previously confirmed transaction with a $(txid,index)$ pair. The witness encodes \emph{input arguments} which \emph{satisfy} the script of the output it spends. 
%---------------------------
%---------------------------

An output is described with the attribute tuple $(out.\textsf{value}, out.\textsf{script})$. The difference between the sum of the output values of a transaction and the value sum of the output referenced by the transaction inputs is denoted the $fee$ which is rewarded to the miner of the block. An output script is a string of operators and is only executed during the validation of a spending transaction, with the spending input \emph{witness} providing the input arguments to the script execution. This is detailed in the following paragraph.

\paragraph{Spending of Outputs} Transactions in Bitcoin are intended to \emph{transfer} the ownership of output values from one participant to another which requires a global ordering of all transactions. Although transactions are grouped in "blocks" in the blockchain, the ordering of all transactions in the blockchain is total. An output can only be redeemed or "spent" by a single input of a transaction ordered thereafter. An output from a confirmed transaction which is not yet redeemed is denoted an unspent transaction output, or UTXO.

In order for a transaction to be valid, all its inputs must contain \emph{witnesses} which satisfy conditions expressed in the output script being spent. When a newly broadcast transaction is \emph{verified}, all scripts of each output it redeems are executed against the respective input witnesses to ensure correct satisfaction. A successful spend of an output requires the script to complete without run-time errors and the script stack to contain an non-zero value stored in the top stack element.

An illustrative output script ($in.prevout.script$) and input witness ($in.witness$) pair are shown below. Note that the witness is copied element-by-element (${w_0, w_1, ..., w_m}$) into an in-memory execution stack ($stack_{init}$) before the script validation run is performed.
\[
\begin{split}
script = [ \: & OP_0, \\
& OP_1, \\
& ..., \\
& OP_n \; ] 
\end{split}
\quad
\begin{split}
witness,\;stack_{init} = [ \: & w_0, \\
& w_1, \\
& ..., \\
& w_m \; ]
\end{split}
\]
The script is executed opcode-by-opcode beginning with its first operator $OP_0$ and $stack_{init}$ until $OP_n$ has been completed. Operator types can read, write or delete the upper region of the script stack. Alternatively, some operator types will perform control flow. Once $OP_n$ is completed, the resulting top stack element must be non-zero for the script evaluation to be successful, thereby validating the input witness under consideration. 

\paragraph{Transaction Time-locks} In aggregate, the transaction time-locks at the input and transaction levels result in an \emph{earliest} possible broadcast time for a given transaction. The earliest broadcast time is the blockheight at which all after and older time-locks in a transaction have released and the transaction can be broadcast and amended to the blockchain.

\emph{Older} time-locks are encoded at the transaction level, and encode an absolute time until which the transaction is time-locked. For example, transaction $A$ in figure \ref{fig:tx_anatomy} is time-locked until blockheight 45. 

\emph{After} time-locks are encoded in each transaction input, which enforces a minimum age of the output being spent. In figure \ref{fig:tx_anatomy}, for example, input 0 of transaction $C$ encodes an \emph{after} time-lock of 2 blocks, which requires transaction $A$ to be at least 2 blocks old when transaction $C$ is broadcast: This time-lock releases at the blockheight $46 + 2 = 48$, since transaction $A$ was amended in block $46$. 

The \emph{earliest} firing time of a transaction occurs when all \emph{after} and \emph{older} time-locks have released. The \emph{After} time-lock of transaction $C$ in figure \ref{fig:tx_anatomy} releases at blockheight 49, and the \emph{older} time-locks at 48 and 50 respectively, given the age of the previous outputs. As a result, the effective earliest broadcast time of transaction $C$ is blockheight 50.

%-------------------------------------------------------------------------------
\subsection{Symbolic Script Execution} \label{Symbolic Execution of Transactions}
%-------------------------------------------------------------------------------

In order to establish an executable, symbolic model of contract transactions, we must be able to perform a symbolic execution of each transaction output script. The intended purpose of an output script is to encode spending conditions, which must be satisfied by the witness of a valid transaction input. A more detailed description of Bitcoin script instruction set semantics can be found in \cite{bitcoinscriptwiki} or in the Bitcoin Core source code \cite{corestagingtree}. Most importantly, a Bitcoin script encodes a finite set of execution paths $\gamma = \gamma_{0} \vee ...\vee \gamma_{n}$. Each output execution path implies a unique set of constraints on the spending input witness, which can be obtained from the symbolic execution of the output script. We have previously noted that symbolic execution is only possible for a fragment of \emph{Script} \cite{klomp2018symbolic} and remains an open challenge. Nonetheless, we illustrate symbolic script execution with Miniscript in section \ref{Miniscript} which relies on an informal definition of the underlying Bitcoin script fragment semantics.

\begin{definition}{\textbf{(Symbolic Witness)}} \label{def:Symbolic Witness}
A symbolic witness $\Gamma$ is a \emph{unique} set of constraints imposed on the individual elements of the spending input witness by an output script. Any input witness which satisfies $\Gamma$ is a valid input witness.
\end{definition}

The following relationship between an input witness $w$ and a symbolic witness $\Gamma$ holds true if the witness satisfies the constraints imposed by the symbolic witness.
\[
w \models \Gamma
\]
Furthermore, we can define a symbolic execution function $sat(output)$ which returns a set of symbolic witnesses $\bar\Gamma$ containing the $n$ alternative script exection paths $\gamma_{0} \vee ...\vee \gamma_{n}$ of the output \emph{script}. 
\[
sat(output) = \bar\Gamma = \{\Gamma_0,..., \Gamma_n\}
\]
For simplicity, we denote $sat(in)$ to mean $sat(in.prevout)$, which returns the symbolic witnesses of the output being spent by $in$. In order to express symbolic witnesses, we adapt and simplify the constraint expressions introduced in \cite{klomp2018symbolic}. In particular, we wish to describe an input witness stack ($w_0, w_1, ..., w_n$) in terms of constraints imposed on each of its elements. The following set $C$ of constraint expressions consists of symbolic witness stack elements $w_{j}$, constant byte strings $b$, constant integers $i$, relational operators and functional expressions such as signature, hash and size:
\begin{equation} \label{eq: C constraints}
\begin{split}
exp \: ::= \: & exp \wedge exp \;| \; \textrm{sig} \; w_{j} \; pk \; | \; \textrm{hash} \; w_{j} = b_{32,20} \; | \; size \; w_{j} = i \; | \\
& w_{j} = b \; | \; after \: >= i  \; | \; older \: >= i
\end{split}
\end{equation}
The symbolic expression $sig \: w_{0} \: pk_{A}$, for example, constrains the first witness stack element ($w_{0}$) to be a valid signature of the transaction signed by the private key $inv(pk_{A})$. The expression $hash \: w_{1} = b_{32}$ constrains the second witness stack element to be the hash pre-image of the 32-byte constant $b_{32}$. The time-lock constraint $older>=i$ in $C$ constrains the value of the $in.\textsf{older}$ attribute and not the witness element directly\footnote{Since both $older$ and $after$ attributes are committed to transaction signatures, of which each transaction input witness generally requires at least one, it can be argued that time-lock constraints $older>=i$ and $after>=i$ are witness constraints on signatures.}. The $after>=i$ time-lock constraint is imposed on the spending transactions $tx.\textsf{after}$ attribute. Example \ref{ex:atomic swap with miniscript} illustrates a symbolic witness composed of symbolic expressions from $C$ for a hash time-lock contract. Our simplified set of constraint expressions can only express symbolic execution of a subset of Bitcoin $Script$, but it is nonetheless sufficiently expressive to describe symbolic execution of Miniscript, which in turn can implement many commonly deployed Bitcoin contracting protocols \cite{herlihy2018atomic} \cite{maxwell2013coinjoin} \cite{poon16bitcoin},  \cite{aumayr2020generalized}. 

We proceed to denote several definitions which can be derived from the analysis of symbolic witnesses. Firstly, if for a specific output execution path an actor can exclusively generate a witness which satisfies the respective symbolic witness, this output path is owned by the actor. 

\begin{definition}{\textbf{(Output Path Ownership)}} \label{def:Output Ownership}
An output path is owned by an actor if $\Gamma \in sat(output)$ and it only features constraints which the actor can exclusively satisfy.
\end{definition}

Such an execution path can be spent anytime by its owner. Of course, the same output may feature alternative execution paths spendable by other actors and thus have multiple owners. If the actor can only exclusively produce the \emph{signatures} of a given output execution path, but does not have the required knowledge to fulfill other non-signature witness constraints, it won't be able to move the funds. However, this actor still maintains control over where the funds can be send for this given output execution path, as the destination outputs are committed by the actor signatures alone. An actor owns an output if it owns all output execution paths.

\begin{definition}{\textbf{(Output Path Control)}} \label{def:Output Control}
An output execution path is controlled by an actor, if it can exclusively satisfy all signature expressions in the respective symbolic witnesses in $\Gamma \in \bar\Gamma$: An actor controlling an output path controls the destination of the output funds when spending along that specific output execution path.
\end{definition}

We now consider the implications of a transaction spending multiple unspent outputs which each feature multiple, alternative symbolic witnesses ($|\bar\Gamma_{i}| > 1$). The spending transaction needs only to satisfy one symbolic witness per input , resulting in a combinatorial expansion of valid, unique witness sets for the spending transaction.

\begin{definition}{\textbf{(Symbolic Witness Permutation)}} \label{def: Symbolic Witness Permutation}
Given a transaction and its inputs ($in_0$, $in_1$,...), each with a possible set of symbolic witnesses $sat(in_i) = \bar\Gamma_i = \{ \Gamma_{0,i}, \Gamma_{1,i}, ... \}$, multiple permutations of valid witness sets are possible over this multiset of symbolic witnesses. A symbolic witness permutation $\pi_i$ for a transaction $tx$ is a unique combination of symbolic witnesses for each tx input. Notation: $\pi(tx)=\{\pi_0, \pi_1, ...\}$ denotes all possible symbolic witness permutations for a given transaction $tx$.
\end{definition}

We conclude this section by illustrating the different script execution paths of a commonly used script pattern called the hash time-locked contract (HTLC).

\begin{example}{\textbf{(Hash Time-locked Contract)}} \label{ex: Hash time-locked Contract}
The following Bitcoin $Script$ example has two different spending conditions $\gamma_{0} \vee \gamma_{1}$, where $\gamma_{0}$ is encumbered with a hash-lock and public key $pk_A$ and $\gamma_{0}$ is encumbered with a delay of 10 blocks and the public key $pk_B$. This script type is commonly referred to as a Hash Time-locked Contract (HTLC), denoting the two script execution paths. 
\end{example}

\begin{lstlisting}
<PK_A> OP_CHECKSIG 
OP_NOTIF
  <PK_B> OP_CHECKSIGVERIFY 
  <10> OP_CHECKSEQUENCEVERIFY
OP_ELSE
  OP_SIZE <32> OP_EQUALVERIFY 
  OP_SHA256 <B_32> OP_EQUAL
OP_ENDIF
\end{lstlisting}

The first execution path $\gamma_{0}$ requires a transaction signature created with the private key $inv(pk_{A})$ (line 1) and the $sha256$ preimage to the digest $b_{32}$ (lines 6,7). Alternative execution path $\gamma_{1}$ can be spent with a transaction signature with the private $inv(pk_{B})$ (line 3) and a time delay of 10 blocks (line 4) after the transaction featuring this script has been amended to the blockchain. Symbolic witnesses $\Gamma_{0}$ and $\Gamma_{1}$ for $\gamma_{0}$ and $\gamma_{1}$ of this \emph{HTLC} are expressed below with constraint expressions (\ref{eq: C constraints})  from $C$ as follows:
\[
\begin{split}
\Gamma_{0} := [ \: & \textrm{sha256} \; w_{0} = b_{32}, \\
& \textrm{sig} \; w_{1} \; pk_{A} \; ] \\
\end{split}
\quad
\begin{split}
\Gamma_{1} := [ \: & \textrm{sig} \; w_{0} \;pk_{B}, \\
& w_{1} = 0, \\
& after >= 10 \: ]
\end{split}
\]

The symbolic witnesses provide the necessary information for an actor to determine whether it can deduce a valid witness for a Bitcoin $Script$ from its knowledge. For example, an actor with the knowledge of $inv(pk_{A})$, the sha256 preimage of $b_{32}$ and the symbolic witness $\Gamma_{0}$, will be able to produce a two element witness stack which satisfies $\Gamma_{0}$ (or the execution path $\gamma_0$). 

We have now defined symbolic witnesses which satisfy specific output execution paths of a Bitcoin output script and demonstrated this information can be used to generate a valid input witness. As previously noted, there exists no full formalization of the Bitcoin script language to enable symbolic execution of the entire Bitcoin $Script$ language to the best of our knowledge. For a subset of Script, Klomp and Bracciali \cite{klomp2018symbolic} have formalized the semantics of individual script operations, so that the constraints expressions in the symbolic witness can inferred step-wise during the symbolic execution of each script command. Alternatively, we propose to make use of the symbolic analysis framework which Miniscript provides, which covers a useful fragment of Bitcoin $Script$.

%-------------------------------------------------------------------------------
\subsection{Miniscript} \label{Miniscript}
%-------------------------------------------------------------------------------

\begin{table*}[h]
  \centering
\small
\begin{tabular}{c|c|c|c|c}
 Miniscript term $m$ & $sat(m)$ & $dsat(m)$ & $script(m)$ & $type(m)$ \\
  \hline
  \hline
  $andor(m_{x}, m_{y}, m_{z})$ & 
    \begin{tabular}{c}
        $sat(m_{y})+sat(m_{x}) \; \vee$ \\ 
        $sat(m_{z})+dsat(m_{x})$ 
    \end{tabular} &
    \begin{tabular}{c}
    % dsatZ+dsatX OR dsatY+satX
        $dsat(m_{z})+sat(m_{x}) \; \vee$ \\ 
        $dsat(m_{y})+sat(m_{x})$ 
    \end{tabular} &
    \begin{tabular}{c}
        $script(ms_x) \; NOTIF \; script(ms_z)$ ...\\ 
        $ELSE \; script(ms_y) \; ENDIF$ 
    \end{tabular} &
    $type(m_{y})$ \\
  \hline
  $and_v(m_{x}, m_{y})$ & $sat(m_{y})+sat(m_{x})$ & $dsat_{y}+sat_{x}$ & $script(m_x) \; script(m_y)$ & $type(m_{y})$ \\
  \hline
  $v(m)$ & $sat(m_{x})$ & - & $script(m_{x}) \; VERIFY$ & V \\
  \hline
  $pk(b_{key})$ & $[sig \; X_{0} \; pk]$ & $[X_{0} = 0]$ & $<b_{key}> \; CHECKSIG$ & B \\
  \hline
  $sha256(b_{32})$ & $[sha256 \; X_{0} = b_{32}]$ & 
     \begin{tabular}{c}
        $[size \; X_{0} = 32 \; \wedge$  \\ 
        $sha256 \; X_{0} != b_{32}]$ 
    \end{tabular} & 
    \begin{tabular}{c}
        $SIZE \; <32> \; EQUALVERIFY$ ...\\ 
        $SHA256 \; <b_{32}> \; EQUAL$ 
    \end{tabular} & 
    B \\
  \hline
  $older(i)$ & $[older >= i]$ & $[older < i]$ & 
  $<i> \; CHECKSEQUENCEVERIFY$ & B \\
  \hline
\end{tabular} \\
\caption{An overview of selected Miniscript terms required to express the HTLC script template from example \ref{ex: Hash time-locked Contract}. Columns $sat(m)$ and $dsat(m)$ denote satisfying and dissatisfying symbolic witnesses for the respective Miniscript term $m$. Both satisfying and dissatisfying symbolic witnesses of child expressions are sometimes necessary for the satisfying witness(es) of the parent Miniscript term.}
\label{table:miniscript}
\end{table*}

\normalsize
This introductory section on Miniscript is not required for the comprehension of Trace-Net and can be skipped by the reader, but we include it nonetheless as Miniscript provides a useful framework for symbolic execution of the underlying Script fragments, with which a Trace-Net model can be instantiated. 

Terms in the Miniscript language $M$ each express a unique $Script$ template, consisting of one or more $Script$ operations. Terms are composable according to Miniscript semantics to produce additional expressions, as shown in the first column of table \ref{table:miniscript}. The semantics of the underlying Bitcoin $script(m)$ of each Miniscript term $m$ are captured by $type(m)$ and type modifier properties. We refer to \cite{wuille2019miniscript} for a full description of Miniscript and focus on a selected subset to illustrate its relevance to Trace-Net. The $script(m)$ column of table \ref{table:miniscript} includes the $Script$ fragments required to parse the raw Bitcoin $Script$ from example \ref{ex: Hash time-locked Contract} into the following Miniscript expression:
\begin{equation} \label{eq: miniscript htlc}
\begin{split}
andor(& \\
& pk_{A}, \\
& sha256(b_{32}), \\
& and_{v}(v(pk_{B}),older(10))\\
& ) \\ 
\end{split}
\end{equation}

Informally, the Miniscript term $andor(m_x, m_y, m_z)$ implies satisfaction expressed as $(m_{x} \wedge m_{y}) \vee m_z$. The Miniscript term $and_v(m_x, m_y)$ requires the satisfaction of both subterms ($m_{x} \wedge m_{y}$). Miniscript distinguishes between satisfying and dissatisfying symbolic witnesses for a given Miniscript term $m$: A dissatisfying symbolic witness provides an input to a $Script$ fragment which executes without a run-time error but does not satisfy its specific execution path. A dissatisfying witness of a Miniscript term is required for the composition of Miniscript parent terms, where the satisfaction or dissatisfaction of a child term may be required for the satisfaction of the parent term. Importantly, however, the satisfying witnesses of the parent Miniscript term can be expressed in terms of the (dis)satisfying witnesses of its child terms, as shown in table \ref{table:miniscript}.

We now illustrate how to construct symbolic witnesses for our HTLC Miniscript (equation \ref{eq: miniscript htlc}) with Miniscript semantics shown in table \ref{table:miniscript}. (Dis)satisfying symoblic witnesses for terminal Miniscript subexpressions such as $pk_{A,B}$, $sha256(b_{32})$ and $older(10)$ are derived first:
\[
\begin{split}
    sat(pk_{A,B}) &= [sig \; w_{0} \; pk_{A,B}] \\
    dsat(pk_{A,B}) &= [w_{0} = 0] \\
    sat(sha256(b_{32})) &= [sha256 \; w_{0} = b_{32}] \\
    sat(older(10)) &= [older >= i]     
\end{split}
\]
At the next expression level, we derive the satisfying symbolic witnesses of $v(pk_B)$ and subsequently $and_{v}(v(pk_{B}), older(10)$ by concatenation of the symbolic witnesses of the child terms according to the $(d)sat(m)$ columns of table \ref{table:miniscript}.
\[
\begin{split}\notag
    sat(v(pk_{B})) &= sat(pk_{B}) = [sig \; X_{0} \; pk_{A,B}] \\
    sat(and_{v}(v(pk_{B}), older(10)) &= sat(older(10))+sat(v(pk_{B})) \\
    &= [sig \; w_{0} \; pk_{B}, \; older >= 10]
\end{split}
\]

Finally, we can infer the symbolic witnesses $\Gamma_{0}$ and $\Gamma_{1}$ of the top-level $andor$ Minsicript expression representing our HTLC script.

\[
\begin{split}
sat(andor(pk_{A},\:sha256(b_{32}&),\;and_{v}(v(pk_{B}),\:older(10))) \\
= sat(sha256(b_{32}&))+sat(pk_{A}) \; \vee \\ sat(and_{v}(v(pk_{B}),\:old&er(10)))+dsat(pk_{A}) \\
 = [sha256 \; w_{0} = b&_{32}, \; sig \; w_{1} \; pk_{A}] \; \vee \\
[sig \; w_{0} \;pk_{B}, \; w_{1} =& \; 0, \; after >= 10] \; \\
= \Gamma_{0} \; &\vee \; \Gamma_{1}
\end{split}
\]

We also note that Miniscript terms are typed and feature type modifier properties, which capture the stack manipulation behavior of the underlying $Script$ fragments, to ensure correct composition of terms: The Miniscript type $B$ is called a \textit{base} expression, as it pushes a non-zero stack element to the stack upon satisfaction. A $V$ type, in contrast, does not add a stack element upon satisfaction, which can be nonetheless useful as a conditional expression in a parent Miniscript term. Further details about the type and properties of Miniscript are defined in \cite{wuille2019miniscript}.

%---------------------------
\begin{figure}[!h]
\includegraphics[width=0.47\textwidth]{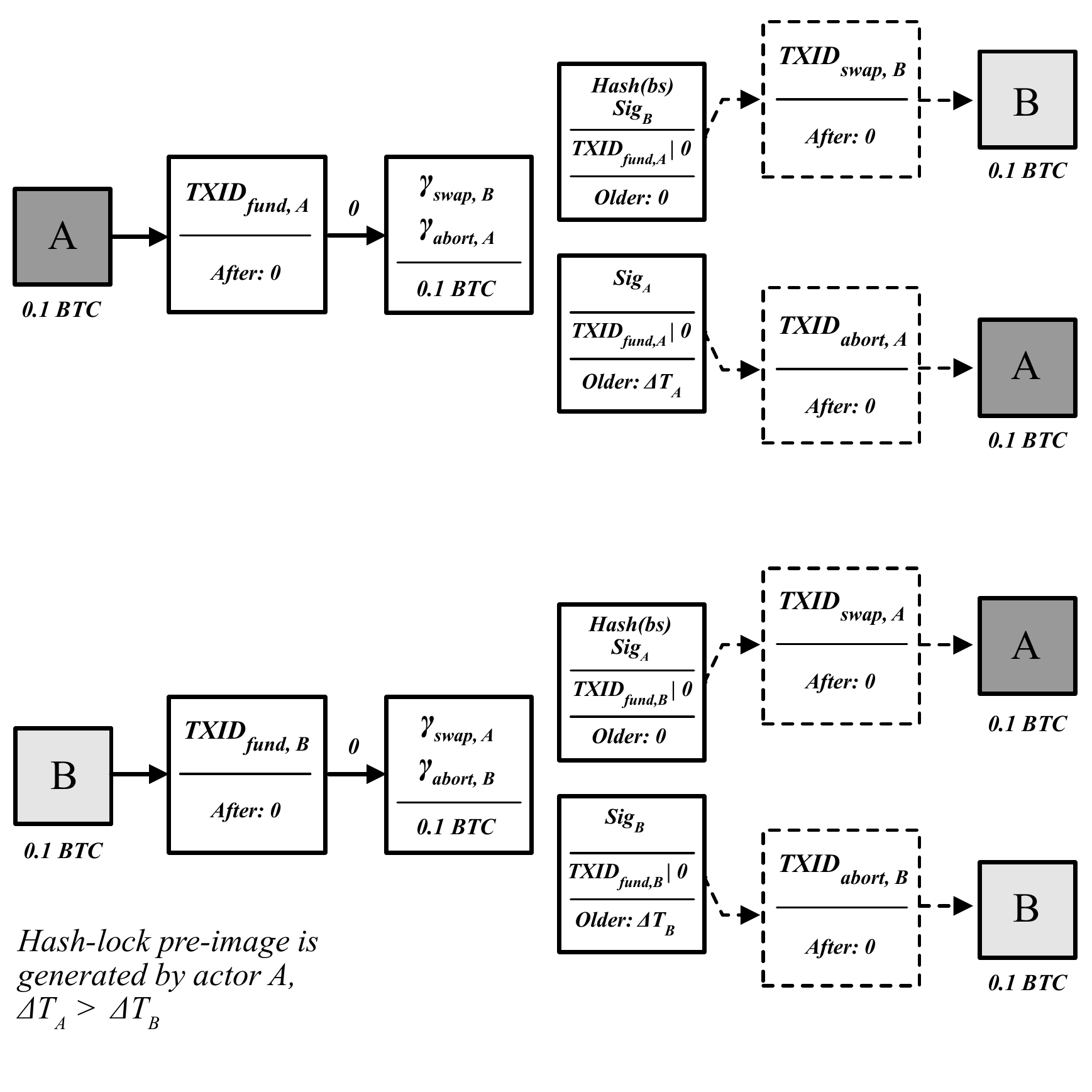}
\caption{A set of contract transactions are shown which stipulate an atomic swap contract between participants A and B. Both transaction outputs feature alternative $\gamma_{swap}$ and $\gamma_{abort}$ execution paths, whilst the identical hash-locks and staggered time-locks enforce the exclusive execution of either swap or abort paths for both parties.}
\label{fig:atomicswaptx}
\end{figure}
%---------------------------

\begin{example}{\textbf{(Atomic Swap Contract with Miniscript)}} \label{ex:atomic swap with miniscript}
We introduce a contract protocol called an atomic swap for which we will step-wise instantiate the Trace-Net model for illustration purposes. The atomic swap protocol is intended for two actors to exchange ownership of Bitcoin outputs. It features HTLC scripts previously introduced in example \ref{ex: Hash time-locked Contract}.
\end{example}

A possible implementation of an atomic swap is shown in figure \ref{fig:atomicswaptx}. Its execution resembles a commit-protocol between the two actors: First, Alice commits her coin to the contract. If Bob responds in like, the coin swap can take place. If Bob does not respond, Alice can withdraw her funds after the expiration of a time-out. In our example implementation, each funding contract transaction features a single output with an HTLC. Both HTLC's feature an identical hash-lock, which functions as a proof of publication\footnote{Once Alice executes her swap transaction, the proof of publication is released on-chain, allowing Bob to execute his swap transaction as well.}. In Miniscript terms, these two output scripts are expressed as follows:  
\[
\begin{split}
HTLC_A=andor(&pk_{B}, sha256(b_{32}), \\
& and_{v}(v(pk_{A}),older(15))) \\ 
\end{split}
\]
\[
\begin{split}
HTLC_B=andor(& pk_{A}, sha256(b_{32}),  \\
& and_{v}(v(pk_{B}),older(10))) \\ 
\end{split}
\]
For both scripts, let us denote the script execution path featuring the hashlocks as the \textit{swap} paths, because they finalize the swap between participating actors. These output paths featuring $pk_A$ or $pk_B$ are controlled by actors A and B respectively, and the private keys for signing remain secret. Swap paths are coupled by the shared hash-lock with a $b_{32}$ secret pre-image generated by a the initiating actor A. Subsequent paragraphs demonstrate how the hash-lock can provide a verifiable publication proof for the swap transaction of the initiating actor. This is required to ensure either both or none of the swap paths are executed.

\textit{Contract Security}\;\;Should the initiating actor A attempt to obtain its swapped funds after both actors have committed funds, the pre-image of this hash-lock will be revealed to actor B upon its broadcast to the Bitcoin network, as a witness element of the swap transaction. Having observed the hash-lock preimage, actor B can now proceed to redeem its swap transaction, thereby completing the contract. The hash-lock therefore functions as a \emph{proof-of-publication}, which performs the "atomic" coupling between the two swap transactions. Omitting the distinction between transaction broadcasts and confirmations, a successful atomic swap initiated by Alice has the following execution trace:
\[
\xrightarrow{tx_{fund,A}}
\xrightarrow{tx_{fund,B}}
\xrightarrow{tx_{swap,A}}
\xrightarrow{tx_{swap,B}}
\circ
\]
Should the Alice become unresponsive after $\xrightarrow{tx_{fund,A}}
\xrightarrow{tx_{fund,B}}$, actor B can wait until the time-lock on its funding transaction output expires, releasing returning actor B's committed funds. Subsequently, with an additional delay, actor A too can regain ownership of its original funds in this abort scenario.
\[
\xrightarrow{tx_{fund,A}}
\xrightarrow{tx_{fund,B}}
\xrightarrow{d_{10}}
\xrightarrow{tx_{abort,B}}
\xrightarrow{d_{5}}
\xrightarrow{tx_{abort,A}}
\circ
\]
Note however, that unsafe contract execution traces are possible with this implementation. Consider the following trace, for example:
\[
\xrightarrow{tx_{fund,B}}
\xrightarrow{tx_{swap,A}}
\circ
\]
If Bob funds the contract first, its funds can be swept by actor Alice who generated the secret pre-image and can execute the swap path of $HTLC_{B}$. Another subtle contract aspect are the delays ($\Delta T_{A}$, $\Delta T_{B}$) imposed on abort transactions $tx_{abort, A}$, $tx_{abort, B}$. For a swap initiated by Alice, who generates the secret pre-image, setting $\Delta T_{A}  = \Delta T_{B}$ will enable the following contract execution trace:
\[
\xrightarrow{tx_{fund,A}}
\xrightarrow{tx_{fund,B}}
\xrightarrow{d(\Delta T_{A/B})}
\xrightarrow{tx_{abort,A}}
\xrightarrow{tx_{swap,A}}
\circ
\]
This trace results in actor A obtaining ownership of both utxo's after executing both $tx_{swap,\:A}$ and $tx_{abort,\:A}$, which is clearly not the intent of the contract design. This unsafe contract outcome is possible, since the time-locks can release simultaneously, enabling race conditions between swap and abort paths: $tx_{swap,A}$/$tx_{abort, B}$ and $tx_{swap,B}$/ $tx_{abort, A}$. In the trace above, we assume the adversary will confirm its transactions first and therefore win both races. A safe contract must therefore be safe despite the possibility of competing transactions. We can solve this by ensuring that $\Delta T_{A}  > \Delta T_{B}$ for a swap contract initiated by Alice, who generates the secret hash pre-image and must confirm a funding transaction first.

Having now considered several trace examples including some which represent unintended contract executions, we proceed to formalize transaction confirmation semantics, which describe how transactions are appended to the blockchain in the Trace-Net model.

%-------------------------------------------------------------------------------
\subsection{Transaction Confirmation Semantics} \label{Transaction Confirmation Semantics}
%-------------------------------------------------------------------------------

\paragraph{Transaction fees} Trace-Net neglects transaction fees, and assumes that these can be sufficiently adjusted without resigning after a transaction is broadcast according to client implementation logics to ensure a transaction is appended to the blockchain within a fixed duration.

\paragraph{Transaction confirmation delays} We differentiate between transactions broadcast by the verifying actor and all other, potentially adversarial external actors. 
The finite confirmation delays for the verifying internal and external actors can be parameterized independently during contract verification (section \ref{section: Automated Trace-Net Analysis}) to accurately reflect contract deployment environments.

\paragraph{Adversarial Blockchain Reorganization} Given that smaller blockchains are often secured by minimal proof-of-work \cite{judmayer2019pay} \cite{moroz2020double}, the cost of performing a 51\% attack and reorganizing the blockchain can be low enough for attackers to pursue. A reorganization allows the attacker to selectively replace and inject transactions in the new chain branch and re-execute the ongoing contract to its advantage. Trace-Net models an adversarial blockchain reorganization at each contract state, to ensure a smart contract safety property will hold despite bounded reorganizations by an adversary (section \ref{section: Automated Trace-Net Analysis}).

%-------------------------------------------------------------------------------
\section{Trace-Net: A Symbolic Model of Contracts} \label{Trace-Net: A Symbolic Model of Contracts}
%-------------------------------------------------------------------------------

Having covered the symbolic execution of Bitcoin transactions, we can now proceed to detail how Trace-Net provides a symbolic model of the entire contract execution. In addition to symbolic analysis of individual output scripts, we must provide a stateful contract model which captures both actor knowledge and blockchain state relevant to the contract. This symbolic contract model must be executable, featuring direct messages exchanges, and on-chain events. We provide an overview of the symbolic Trace-Net Model before defining state transition semantics in subsequent subsections.

\paragraph{Internal \& External Actors} In our model of Bitcoin contract protocols, each verifying actor will assume that all other counter-parties are colluding against the verifying actor. We can thereby simplify each contracting protocol as one between two actors: The internal, verifying actor and the adversarial, external actor. This implies that direct contract protocol messaging can be modelled as single bi-directional communication channel between these two actors, with no additional relevant intruders. 

\paragraph{Actor Knowledge} We model the actor knowledge states as $K_{int},K_{ext}$, consisting of symbolic knowledge objects such as transaction templates and secrets. Actors always expand their knowledge when it is updated with a direct message from the counter-party or an observation on the blockchain. Knowledge expansion is performed with a set of public functions $\Sigma$ available to all actors, modeling Dolev-Yao deductive abilities of an actor during contract protocol execution (section \ref{Actor Knowledge}).

\paragraph{Setup phase} Each contract protocol implementation begins with the generation of the contract transaction templates during the \emph{setup} phase. Transaction templates have fully defined attributes, but feature empty witnesses, and can therefore not be broadcast. We denote a transaction template as $tx^{\prime}$ and a completed, valid contract transaction with valid witnesses as $tx$. Informally, the setup phase corresponds to the $stipulation$ of the contract for the participating actors. The setup involves no exchanges of signatures or generated secrets. We express the generation of the contract transaction templates as a implementation-defined function $\textsf{GenTx}$, which, on an input of public keys $PK^{*}$, hash digests $B^{*}$ and funding output sets $outs_f^{*}$, will output a set of contract transaction templates $TX^{\prime}$. These required inputs to $\textsf{GenTx}$ are defined by the contract implementation, but require at least one public key and one funding output in total.
\[
\textsf{GenTx}(PK^{int},PK^{ext}, B^{int}, B^{ext}, outs_f^{int}, outs_f^{ext}) = TX^{\prime}
\]
 Inputs to and output of $\textsf{GenTx}$ are known by both internal and external actors at the end of the setup phase. Actors have access to the following public functions in $\Sigma_{setup}$ during the setup phase:
\begin{equation} \label{eqn:setup}
\Sigma_{setup}=(\textsf{Gen},\:\textsf{PK},\:\textsf{Hash},\:\textsf{GenTx}) 
\end{equation}
The function $\textsf{Gen}$ is a from a Bitcoin supported signature scheme $\Xi = (\textsf{Gen}, \textsf{Sign}, \textsf{Vrfy})$, $\textsf{PK}$  the public key derivation function and $\textsf{Hash}$ a hash function supported by Bitcoin script. For simplicity, actors call \textsf{Gen}
to generate both private key or hash pre-image secrets required for the public key and hash inputs of \textsf{GenTx}. These secrets are added to the actors knowledge during the setup phase: Once this is complete, Trace-Net can be instantiated to symbolically execute the contract and generate the reachable state-space for safety verification.

\paragraph{Symbolic execution phase} Trace-Net models a symbolic contract state with the tuple $(K_{int}, K_{ext}, B)$, where $K$ is the state of an actor's knowledge and $B$ the on-chain state of the contract Bitcoin transactions. In each protocol state during the execution phase, the following transaction types can be \emph{fired} by the actors if the transitions are in a \emph{fireable} state.

\begin{itemize}
  \item $\textit{e}$ - Message exchange 
  \item $\textit{tb}$ - Transaction broadcast 
  \item $\textit{t}$ - On-chain confirmation
  \item $\textit{d}$ - Time delay 
  \item $\textit{r}$ - Blockchain reorganization
\end{itemize}

We provide an initial introduction to the transition types and formally define both transition firing semantics in later sections. For actor $a,b \in \{int,ext\}$:

$(K_a,K_b,B) \xrightarrow{e^{a}} (K_a,K_b^{\prime},B)$: A message exchange transition is a direct message between internal and external actors, where the knowledge of recipient is updated with the message object. A message exchange transition only updates the knowledge state of the receiving actor (definitions \ref{def: Fireable Message Transition}, \ref{def: Message Transition Firing}).

$(K_a,K_b,B) \xrightarrow{tb^{a}} (K_a,K_b^{\prime},B)$: When an actor intends to add a valid transaction to the Bitcoin blockchain, the transaction must first be broadcast to the Bitcoin peer-to-peer network with a transaction broadcast transition, fired by an actor who can produce the transaction from knowledge. A broadcast is only modeled in Trace-Net if it results in an update of the counterparty's knowledge (definitions \ref{def: Fire-able Transaction Broadcast}, \ref{def: Transaction Broadcast Firing}).

$(K_a,K_b,B) \xrightarrow{t^{a}} (K_a,K_b,B^{\prime})$: An on-chain transaction can only be confirmed or amended to the blockchain if it has previously been broadcast or is already fully known by the counterparty. An on-chain transition only updates the on-chain contract state, and can be preceded by a time delay to model the required confirmation time (definitions \ref{def: Fire-able On-chain Transition}, \ref{def: On-chain transition firing}). 

$(K_a,K_b,B) \xrightarrow{d} (K_a,K_b,B^{\prime})$: A time delay interval is expressed in a number amended blocks and like an on-chain transition only affects the on-chain state $B$ (definition \ref{def: time transition}). 

$(K_{int},K_{ext},B) \xrightarrow{r^{ext}(n)} (K_{int}/K_{int}^{\prime},K_{ext},B^{\prime})$: A blockchain reorganization of depth $n$ is fireable by the external actor at any contract state. Since the external actor can freely determine the reorganization depth and transaction ordering of the new chain branch, different reorganizations up to a maximum depth are possible at a given state (definition \ref{def: Chain roll-back transition firing}, section \ref{section: Automated Trace-Net Analysis}).

\paragraph{Actor Strategies} We refer to the set of \emph{fire-able} transitions of an actor as its \emph{strategies} in contract state $(K_{int}, K_{ext}, B)$. 

%-------------------------------------------------------------------------------
\subsection{Actor Knowledge Derivation} \label{Actor Knowledge}
%-------------------------------------------------------------------------------

Actor knowledge is modeled in a Dolev-Yao-fashion \cite{dolev1983security}, where each actor can access a set of public function to deduce additional information during the symbolic execution of the underlying contract implementation.

\begin{definition}{\textbf{(Actor Knowledge)}}
An actor knowledge $K$ is a finite set of transaction templates, private keys, public keys, hash pre-images and signatures. $K$ can be expanded with a set of public functions $\Sigma_{exec}$ during symbolic contract execution whenever $K$ is updated with external information.
\end{definition}

Let $k$ be a knowledge object which is derived from $K$ with a function in $\Sigma_{exec}$. We denote the knowledge expansion of $K$ with $k$ with the following notation. 
\[
K \vdash k
\]
The public functions $\Sigma_{exec}$ allow actors to derive sweep transaction templates from outputs and transaction signatures from private keys.
\begin{equation} \label{eqn:exec}
\Sigma_{exec}=(\textsf{SweepTx}, \textsf{Sign})
\end{equation}

\begin{itemize} 
\itemsep0em 
\item $\textsf{SweepTx}(out)$ is a function that generates a symbolic sweep transaction template spending $out$, which features an execution path \textit{controlled} (definition \ref{def:Output Control}) by a single actor $a \in \{int, ext\}$. This sweep transaction sends the funds to single output owned by actor $a$. 

\begin{prooftree}
\AxiomC{$K \vdash out$}
  \RightLabel{$if\;ctrl(out,actor)$}
  \UnaryInfC{$K \vdash \textsf{SweepTx}(out,actor)$}
\end{prooftree}

Where $ctrl(out,actor)$ denotes that $out$ features an execution path controlled by $actor$. $K \vdash out$ if the output is a member of any transaction template in the knowledge of actor $actor$.

\item $\textsf{Sign}(tx^{\prime},inv(pk))$ is a function that on input transaction template $tx^{\prime}$ and secret key $inv(pk)$ outputs a signature $sig_{pk}(tx^{\prime})$. This signature satisfies the constraint $[sig \; w \; pk]$ as $w$. 
\begin{prooftree}
\AxiomC{$K \vdash tx^{\prime}$}
\AxiomC{$K \vdash inv(pk)$}
  \RightLabel{$if\;pk\in \textsf{sat}(tx^{\prime}.ins)$}
  \BinaryInfC{$K \vdash \textsf{sign}(tx^{\prime},inv(pk))$}
\end{prooftree}

This derivation rule is applicable to all public keys featured in the symbolic witnesses of transaction template inputs ($pk$ in $\textsf{sat}(tx^{\prime}.ins)$).

\end{itemize}

\noindent
\emph{Bounded knowledge expansion} with $\Sigma_{exec}$ implies a finite contract execution state-space. Inputs and outputs of public functions \textsf{SweepTx} and \textsf{Sign} are typed, such that derived knowledge cannot be used as an inputs to $\Sigma_{exec}$. \textsf{SweepTx} consumes \emph{controlled} outputs and produces \emph{owned} outputs. Likewise, signatures generated from the public \textsf{Sign} function cannot be used as inputs for $\Sigma_{exec}$. 

\begin{example}{\textbf{(Initial Knowledge Expansion)}}
In the following example, we illustrate how the actor knowledge is expanded at the beginning of the symbolic execution phase of an atomic swap contract. 
\end{example}

For our \emph{atomic swap} example initiated by the internal actor, the initial knowledge $K^{int}$ of internal actor after the \emph{setup} phase executed by the contract implementation is initialized as follows. Note that we do not explicitly list the funding outputs, as these are already referenced as $in.prevout$ attributes in the funding transactions, which the actor can extract. For simplicity, we assume that actors reuse private keys for all outputs.
\[
\begin{split}
K_{int} = \{ & inv(pk_{int}),b_{32}, tx^{\prime}_{fund, int},tx^{\prime}_{fund, ext}\}
\end{split}
\]
Subsequently, since $tx^{\prime}_{fund, int}$ and $tx^{\prime}_{fund, ext}$ contain the funding outputs as well as swap and abort output execution paths which are all controlled by single actors, we can expand the internal actor's knowledge with sweep transactions (\textsf{SweepTx}) for all these output execution paths. These sweep transactions (sweep, swap, abort) are depicted in figure \ref{fig:atomic_swap_petri.pdf}.
\[
\begin{split}
K_{int} \vdash \{ & tx^{\prime}_{swap, int}, tx^{\prime}_{abort, int}, tx^{\prime}_{sweep, int}, \\
& tx^{\prime}_{swap, ext}, tx^{\prime}_{abort, ext}, tx^{\prime}_{sweep, ext}, \\
& sig_{pk_{int}}(tx^{\prime}_{fund, int}), \;sig_{pk_{int}}(tx^{\prime}_{swap, int}), \\
& sig_{pk_{int}}(tx^{\prime}_{abort, in}),sig_{pk_{int}}(tx^{\prime}_{sweep, int}) 
\}
\end{split}
\]
The internal actor furthermore can derive signatures which are required to satisfy the the symbolic witnesses featuring public key $pk_{int}$ with \textsf{Sign}. The knowledge of the internal actor has now been fully expanded in the initial state of the symbolic contract execution. 

%-------------------------------------------------------------------------------
\subsubsection{Cryptographic Extensions} \label{Cryptographic Extensions}
%-------------------------------------------------------------------------------

It is possible to extend $\Sigma_{setup}$ (eq. \ref{eqn:setup}) and $\Sigma_{exec}$ (eq.\ref{eqn:exec}) to model additional cryptographic sub-protocols useful for the construction of contracts. We demonstrate such an extension with the adaptor signature scheme, previously formalized in \cite{aumayr2020generalized}. An adaptor signature scheme consists of $\Xi_{adapt} = (\textsf{pSign},\:\textsf{pVrfy},\:\textsf{Adapt},\:\textsf{Ext})$ for a signature scheme $\Xi_{sig} = (\textsf{Gen},\:\textsf{Sign},\:\textsf{Vrfy})$. 

\begin{itemize} 
\itemsep0em 
\item $\textsf{pSign}(tx^{\prime},\textsf{inv}(pk),pk_y)$ is a function that on input a message $tx^{\prime}$, private key $inv(pk)$ and public key $pk_y$ outputs a pre-signature $psig_{pk}(tx^{\prime},pk_y)$
\item $\textsf{pVrfy}(psig_{pk}(tx^{\prime},pk_y),tx^{\prime},pk,pk_y)$ is a function that verifies a pre-signature against a message $tx^{\prime}$ and public keys $pk$ and $pk_y$. 
\item $\textsf{Adapt}(psig_{pk}(tx^{\prime},pk_y),inv(pk_y)$ is a function that on input a pre-signature $psig_{pk}(tx^{\prime},pk_y)$ and a private key $inv(pk_y)$ returns a signature $sig_{pk}(tx^{\prime})$.
\item $\textsf{Ext}(psig_{pk}(tx^{\prime},pk_y), sig_{pk}(tx^{\prime}),pk_y$) is a function that on input a pre-signature, corresponding signature and public key $pk_y$ returns the a private key $inv(pk_y)$.
\end{itemize}

Given pre-signature $psig_{pk}(tx^{\prime},pk_y)$, an observer can extract $inv(pk_y)$ upon learning $sig_{pk}(tx^{\prime})$ with the function $\textsf{Ext}$. Conversely, given $psig_{pk}(tx^{\prime},pk_y)$ and $inv(pk_y)$, a valid signature $sig_{pk}(tx^{\prime})$ can be inferred. 

\begin{example}{\textbf{(Atomic Swap with Adaptor Signatures)}}
In the following example, our atomic swap contract has the hash-lock in the swap spending paths replaced with pre-signatures from $\Xi_{adapt}$. The pre-signatures are known by both actors but not observable on-chain.
\end{example}

We adapt the output scripts for atomic swap transactions from example \ref{ex:atomic swap with miniscript} and replace the hash-lock in the \emph{swap} spending condition with a public key of the counterparty. Expressed in Miniscript, the HTLC output scripts of $tx_{fund, in}$ and $tx_{fund,ex}$ for our atomic swap contract initiated by the internal actor are the following:
\[
\begin{split}
m_{int}=andor(&pk_{ext}, pk_{int}, \\
& and_{v}(v(pk_{int}),older(15))) \\ 
\end{split}
\]
\[
\begin{split}
m_{ext}=andor(& pk_{int}, pk_{ext},  \\
& and_{v}(v(pk_{ext}),older(10))) \\ 
\end{split}
\]
\noindent
With the hash-lock removed, there are no on-chain transaction semantics which reveal a coupling between the swap paths of the two contract participants. Instead, this coupling is setup during the user-defined \emph{setup} protocol with an exchange of pre-signatures between the actors. 

\emph{Setup Phase:}\;The user-defined setup phase now features an extended public function set.
\[
    \Sigma_{setup}^{\prime} = \{ \textsf{Gen}, \textsf{PK}, \textsf{Hash}, \textsf{genTx}, \textsf{pSign}, \textsf{pVrfy} \}
\]
\noindent
Note that GenTx must also generate swap transactions for both actors, so that both actors can generate and verify pre-signatures for swap transactions $tx_{swap,int}^{\prime}, tx_{swap,ext}^{\prime}$ of the counter-party. $inv(pk_y)$ is generated by the initiating, internal actor and so $inv(pk_y)\notin K_{ext}$ at setup. Once these pre-signatures are validated, the setup protocol is complete. When compared to the previously illustrated atomic swap contract with hash-locks, the actor knowledge after contract setup are extended by pre-signatures and related public keypairs ($inv(pk_y)$, $pk_y$).
\[ 
\begin{split}
K_{int} = \{ 
& ..., inv(pk_y), \\
& psig_{pk_{ext}}(tx^{\prime}_{swap,int},pk_y), psig_{pk_{int}}(tx^{\prime}_{swap,ext},pk_y)
\}
\end{split}
\]
\[
\begin{split}
K_{ext} = \{ 
& ..., pk_y, \\
& psig_{pk_{int}}(tx^{\prime}_{swap,ext},pk_y), psig_{pk_{ext}}(tx^{\prime}_{swap,int},pk_y)
\}
\end{split}
\]
\noindent
The set of public functions available to actors $\Sigma_{exec}$ during symbolic contract execution is extended with $\textsf{Adapt},\textsf{Ext} \in \Xi_{adapt}$. 
\[
    \Sigma_{exec}^{\prime} = \{ \textsf{SweepTx}, \textsf{Sign},\textsf{Adapt}, \textsf{Ext} \}
\]
We only illustrate the contract state during symbolic execution when both funding transactions are confirmed and no direct messages have been exchanged yet. The initiating internal actor wishes to execute its swap transaction: The spending conditions for the swap path of output script $m_{ext}$ require $sig_{pk_{int}}(tx^{\prime}_{swap, int})$ and $sig_{pk_{ext}}(tx^{\prime}_{swap, int})$, the latter which is derived with the \textsf{Adapt} function.
\small
\begin{prooftree}
\AxiomC{$K_{int} \vdash psig_{pk_{ext}}(tx^{\prime}_{swap,int},pk_y)$}
\AxiomC{$K_{int} \vdash inv(pk_y)$}
    \RightLabel{$\textsf{Adapt}$}
    \BinaryInfC{$K_{int} \vdash sig_{pk_{ext}}(tx^{\prime}_{swap, int})$}
\end{prooftree}

\normalsize
\noindent
Once $tx_{swap,int}$ is executed, the external actor can observe $sig_{pk_int}(tx^{\prime}_{swap, int})$ (publication proof), and derive $inv(pk_y)$ with the \textsf{Ext} function.
\small
\begin{prooftree}
\AxiomC{$K_{ext} \vdash psig_{pk_{ext}}(tx^{\prime}_{swap,int},pk_y)$}
\AxiomC{$K_{ext} \vdash sig_{pk_{ext}}(tx^{\prime}_{swap,int},pk_y)$}
    \RightLabel{$\textsf{Ext}$}
    \BinaryInfC{$K_{ext} \vdash inv(pk_y)$}
\end{prooftree}

\normalsize
\noindent
The external actor can expand its knowledge with \textsf{adapt}, the pre-signature and publication proof to obtain $sig_{pk_{ext}}(tx^{\prime}_{swap,ext},pk_y)$. This signature enables the external actor to also execute its swap transaction, thereby completing the contract.
\small
\begin{prooftree}
\AxiomC{$K_{ext} \vdash psig_{pk_{ext}}(tx^{\prime}_{swap,ext},pk_y)$}
\AxiomC{$K_{ext} \vdash inv(pk_y)$}
    \RightLabel{$\textsf{Adapt}$}
    \BinaryInfC{$K_{ext} \vdash sig_{pk_{int}}(tx^{\prime}_{swap, ext})$}
\end{prooftree}

\normalsize
Note that without the shared hash-lock, the atomic swap scheme with adaptor signature does not reveal any contract coupling between swap transactions on the Bitcoin blockchain: The atomic coin swap contract is now no longer visible to the observer of the Bitcoin network.

%-------------------------------------------------------------------------------
\subsubsection{Messages \& Transaction Broadcasts} \label{Actor Messages & Transaction Broadcasts}
%-------------------------------------------------------------------------------
In addition to deducing information from existing knowledge with public functions, actors can directly exchange messages to update the knowledge of the recipient. Furthermore, a transaction broadcast by an actor which includes information previously unknown to the counter-party also results in the knowledge update of the observer.

\begin{definition}{\rm\textbf{(Fire-able Message Transition)}} \label{def: Fireable Message Transition}
A message transition $e^{a}(k)$ with payload $k$ is fire-able by an actor $a \in \{int,ext\}$ with knowledge $K_a$, if $k$ is known by the sender but unknown to the recipient. 
\end{definition}

\begin{definition}{\rm\textbf{(Message Transition Firing)}} \label{def: Message Transition Firing}
Let $e^{a}(k)$ be a transition for actor knowledges $(K_a,K_b)$, fire-able by actor $a \in \{int,ext\}$. The firing of of $e^{a}(k)$ updates actor knowledge $(K_a,K_b)$ to $(K_a, K_{b}\cup \{k\})$
\end{definition}

\begin{prooftree}
\AxiomC{$K_{a} \vdash k$}
\AxiomC{$K_{b} \cancel{\vdash} k$}
  \BinaryInfC{$K_b \xrightarrow{e^{a}(k)} K_b\;\cup\;\{k\}$}
\end{prooftree}

An actor knowledge can also be expanded with knowledge observed on the blockchain. Specifically, an actor will expand its knowledge when the counter-party broadcasts a transaction on the network and the transaction includes previously unknown input witness elements such as signatures or hash pre-images.

A transaction broadcast is only fireable during the symbolic execution of a contract if it leads to a knowledge update of the observing actor. In order for an actor to fire a broadcast transition, it must be deducible by the initiating actor, be consensus valid according to a blockchain model B, and provide new information to the counterparty. A complete, valid transaction can be deduced by an actor if can produce a valid witness for each of its inputs: A valid witness stack for an input must satisfy at least \emph{one} of the symbolic witnesses $\{\Gamma_0,\Gamma_1,...\}=sat(in)$. The deduction rule for an actor to derive a valid transaction is shown below:

\small
\begin{prooftree}
\AxiomC{$K_{a} \vdash tx^{\prime}$}
% \AxiomC{$(K \vdash w.wit \models \Gamma.\Gamma \in sat(in))\;for\;in\in\;tx.ins$}
\AxiomC{$\forall(in\in tx^{\prime}.ins)\exists(w,\Gamma)(K_{a} \vdash w \wedge w \models \Gamma \wedge \Gamma \in sat(in))$}
    \BinaryInfC{$K_{a} \vdash tx$}
\end{prooftree}

\normalsize
\noindent
The actor must know the transaction template, and be able to generate a valid witness $w$ for each transaction template.

\begin{definition}{\rm\textbf{(Fire-able Transaction Broadcast)}} \label{def: Fire-able Transaction Broadcast}
A transaction broadcast tb for a transaction is fire-able if it is deducible by actor $a \in \{int,ext\}$, is valid in blockchain state B and contains witness stack elements unknown to actor b.
\end{definition}
\begin{definition}{\rm\textbf{(Transaction Broadcast Firing)}} \label{def: Transaction Broadcast Firing}
The firing of a transaction broadcast leads to an update of the observing actors knowledge if it contains witness elements previously unknown to the observer.
\end{definition}

\small
\begin{prooftree} 
\AxiomC{$K_B \vdash tx$}
\AxiomC{$\exists{w_{i}}(w_{i} \in tb^{a}.ins \wedge K_b \cancel{\vdash} w_i)$}
\AxiomC{$tx\;is\;valid\;in\;B$}
    \TrinaryInfC{$K_b \xrightarrow{tb^{a}} K_b\;\cup\;\bigcup_{w_{i}\in tb^{a}.ins\;\wedge\; K_b\cancel{\vdash}\;w_{i}}\{w_{i}\}$}
\end{prooftree}

\normalsize
\noindent
Witness elements of the broadcast transaction $tx$ are denoted as $w_{i} \in tb^{a}.ins$ in the firing rule shown above.

%-------------------------------------------------------------------------------
\subsection{Trace-Net Blockchain Model} \label{Petri Net Output State Model}
%-------------------------------------------------------------------------------

The Trace-Net blockchain model captures the on-chain state of contract transactions. To this end, Trace-Net adopts a classic Petri Net skeleton to capture the state and fire-ability of on-chain transitions based on the availability of unspent outputs and the current height of the Bitcoin blockchain.

\begin{figure}
\begin{center}
\includegraphics[width=0.47\textwidth]{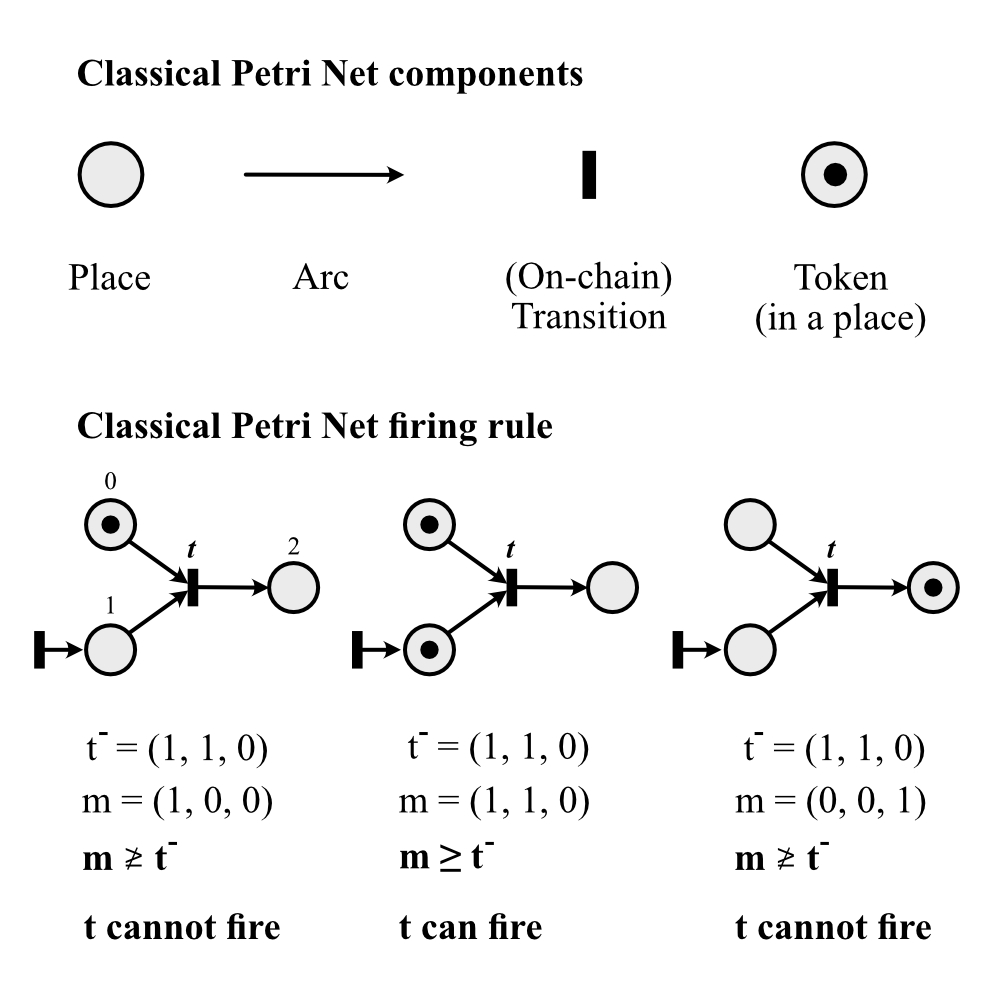}
\end{center}
\caption{ Classical Petri Net components consist of places, arcs, transitions and tokens. The classical Petri Net firing rule denotes that a transition can only fire if its input arcs are connected to places populated with tokens.}
\label{fig:classicpetrinet}
\end{figure}

There are four classical Petri-Net elements \cite{petrinetbook} adopted by the Trace-Net: Places, tokens, arcs and on-chain transitions, as shown in figure \ref{fig:classicpetrinet}. Tokens can only exist in places. Each on-chain transition is connected to at least one place with directed arcs. Arcs pointing from a place to an on-chain transition reflect transaction inputs and are denoted input arcs $(p,t)$. Arcs pointing to places represent transaction outputs and are denoted output arcs $(t,p)$. A given place $p$ can only connect to a single input and single output arc at most and represents a transaction output. 

When an on-chain transition fires, the input arcs of the fired transition each consume a token from their connected place. Simultaneously, each transition output arc will produce a token to its connected place during firing, modeling the consumption and production of unspent transaction outputs on the blockchain during the execution of the contract protocol. The place marking function $m(p) \in \{0,1\}$ returns the number of tokens currently in a place $p$. We extend the Petri Net formalism with time markings and earliest firing time intervals to arrive at the definition of a Trace-Net.

\begin{definition}{\rm\textbf{(Trace-Net)}} 
\label{def: Trace-Net}
A Trace-Net $\mathcal{Z}$ is the tuple $(K^{int}_0, K^{ext}_0,P,T,F, I_{older},I_{after},m_0,b_{0})$ such that
\end{definition}
\begin{enumerate}
	\item $K^{int}_0, K^{ext}_0$ is the initial actor knowledge.
	\item There exists a $p(out) \in P$ for each $out \in K^{in}_0/K^{ex}_0$.
	\item F is a relation $F\subseteq(P \times T) \cup (T \times P)$.
	\item For all $tx \in K^{int/ext}_0$: 
	\begin{enumerate}
    	\item There exists a unique $t \in T$ for each $\pi_i$ of $\pi(tx)$. 
    	\item For each $tx$ and $in \in tx$, $(p(in.prevout),t)\in F$. 
    	\item For each $tx$ and $out \in tx$, $(t,p(out)) \in F$. 
	\end{enumerate}
    \item $I^{after}, I^{older}:(P \times T) \rightarrow  \mathbb{N}^0$
    \begin{enumerate}
        \item $I_{after}(p,t)$, $I_{after}(p,t) \models t.\pi_i$ for $t \in T$
    \end{enumerate}
	\item $m_0$ : $P \rightarrow \{0,1\}$ is the initial marking
	\item $b_0$ : $ \rightarrow \mathbb{N}$ is the initial blockheight
\end{enumerate}

(2) A place is added to $\mathcal{Z}$ for each output featured in the contract transaction templates present in the initial actor knowledge states. This includes outputs referenced by transaction templates (funding outputs) and symbolic sweep transactions. (3) F a relation representing input and output arcs. (4) Trace-Net models a dedicated transition for each symbolic witness permutation (definition \ref{def: Symbolic Witness Permutation}) of a transaction: This enables the fireability of each symbolic witness permutation to be modeled separately (5) Earliest firing intervals is a function over input arcs and must satisfy the symbolic witness permutation associated with its transition. The firing intervals are implied by the $after$ and $older$ constraints expressed in the respective symbolic witness. 

%---------------------------
%---------------------------

Let us denote $t^{-}$ as the required token availability for $t$ over all places $P$ in Trace-Net $\mathcal{Z}$ in order for $t$ to be fireable. This is illustrated in figure \ref{fig:classicpetrinet}. 

\[
t(p)^{-}=\begin{cases}1 & if\; (p,t)\in F\\0 & if \; (p,t)\not\in F\end{cases}
\]
Similarly, we can denote $(p,t)^{-}$ as the required token availability for place $p$ required by an input arc, which will trivially be 1 for a connected place and 0 for a disconnected one.

%---------------------------
\begin{figure*}[ht]
\begin{center}
\includegraphics[width=\textwidth]{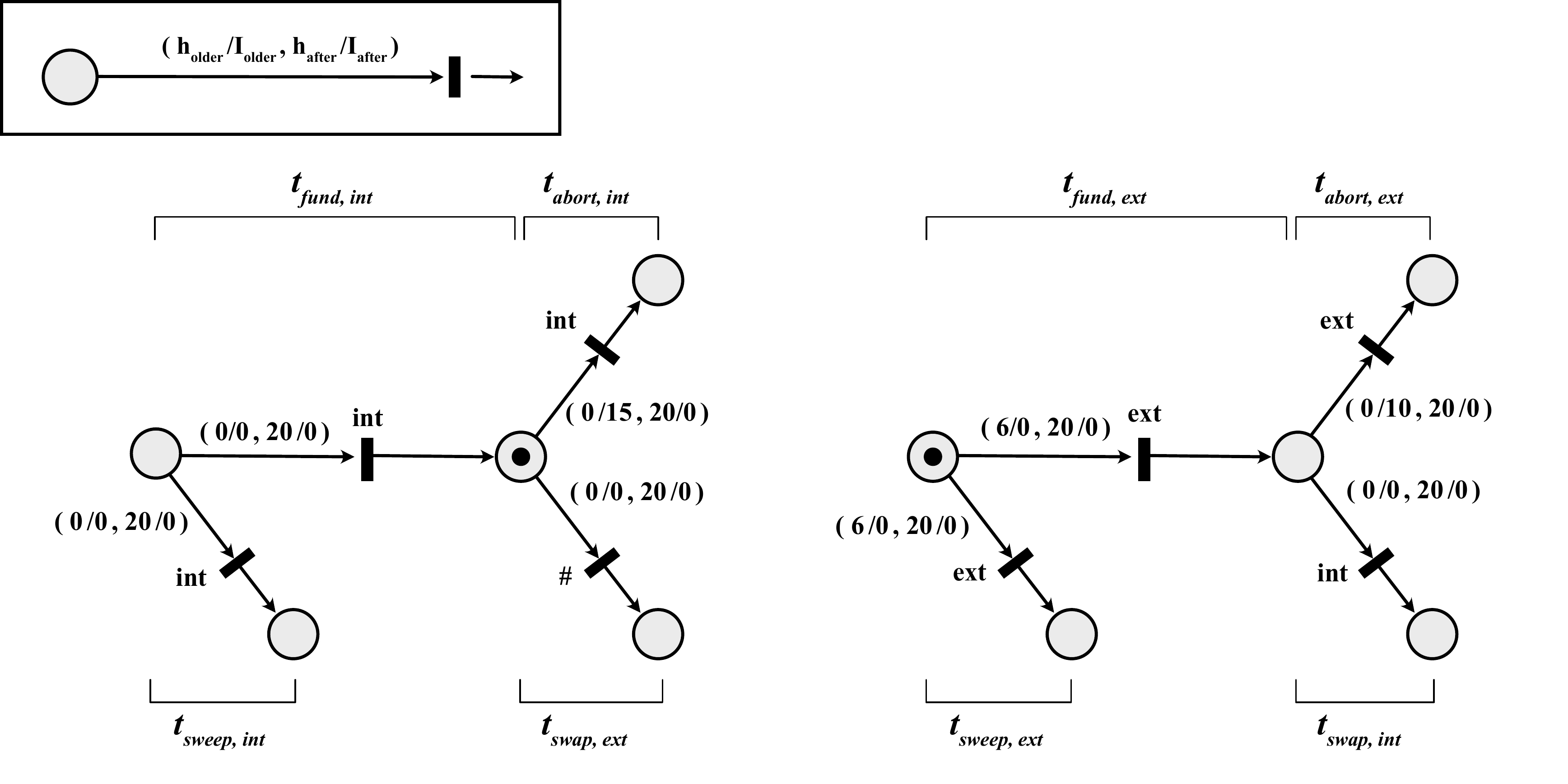} \label{fig:atomic_swap_petri.pdf}
\end{center}
\caption{A Trace-Net model of an atomic swap initiated by the internal actor is shown. Input arc and transition markings are overlayed on top of the Trace-Net skeleton. Transitions are annotated with the actors who can produce the corresponding valid transaction witnesses from knowledge. The state shown above represents the contract state $z$ imminently after the confirmation of the internal actor's funding transaction. Note the chain-height is 20 at time of evaluation, and that abort delays are set to 15 and 10 blocks respectively.}
\label{fig:atomic_swap_net}
\end{figure*}
%---------------------------

Trace-Net $\mathcal{Z}$ models the possible on-chain contract transactions and the possible states of unspent outputs throughout the symbolic execution of the contract. However, in order to model the contract state, we must also introduce a notion of time to determine the expiration of time-locks.

\begin{definition}{\rm\textbf{(Time Markings)}}
 A function $h_{*}:(P \times T) \rightarrow \mathbb{N}^0$ is a time marking in $\mathcal{Z}$.
\end{definition}

In order to express the satisfaction or dissatisfaction of time constraints imposed by spending conditions featuring $(after >=i)$ or $(older >i)$ expressions, stateful time markings are applied to input arcs. For an on-chain transition to be fire-able, all time markings must have released the respective on-chain transition. A time marking releases its input arc when they reach or exceed their respective earliest firing times.

\begin{enumerate}
  \item $h_{older}(p,t) \geq I_{older}(p,t))$
  \item $h_{after}(p,t) \geq I_{after}(p,t))$
\end{enumerate}

We can now define a contract state in the Trace-Net framework. It consists of the state of actor knowledge, time markings, place markings and on-chain history of confirmed blockchain transactions. An example of a contract state is depicted in figure \ref{fig:atomic_swap_net}, which represents the contract state of an atomic swap initiated by the internal actor after the contract has been unilaterally funded by the internal, verifying party. 

\begin{definition}{\rm\textbf{(Trace-Net State)}} \label{def: Trace-Net State}
Let a state in Trace-Net Z be the tuple $z := (K_{in}, K_{ex}, B) = (K_{in}, K_{ex}, h_{older}, h_{after}, m, q)$ such that
\end{definition}

\begin{enumerate}
  \item For all $(p,t)\in F:$
  \begin{enumerate}
    \item $((p,t)^{-}< m)  \rightarrow h_{older}(p,t)=0$
    \item $((p,t)^{-}= m)  \rightarrow h_{older}(p,t) \in \mathbb{N}^0$
    \item $h_{after}(p,t) \rightarrow \mathbb{N}$
  \end{enumerate}
    \item $m$ is the current place marking
    \item $q$ is the current on-chain history
\end{enumerate}

(1) The input arc time marking $h_{older}(p,t)$ can only begin to increment when a symbolic unspent output becomes available. $h_{after}$ is always equal to the blockheight, as it only enforces a global expiration time unrelated to output confirmation ages.

\begin{definition}{\rm\textbf{(On-chain history)}} \label{On-chain chronology}
The on-chain history at a Trace-Net state is the history of the on-chain contract transactions appended to the blockchain.
\end{definition}
On-chain histories can be expressed as a list of tuples $q = \{(t_{0}, h_{0}), (t_{1}, h_{1}),...\}$ ordered according to their position in the blockchain, where $h_i$ represents the blockheight of the transaction corresponding to $t_i$. 

\begin{definition}{\rm\textbf{(Time transition d)}} \label{def: time transition}
Let d be a non-zero natural number, and z a state in $\mathcal{Z}$, then the elapsing of time d will change state z into $z^{\prime}$ such that
\end{definition}

\begin{enumerate}
  \item $K_{in}^{\prime}, K_{ex}^{\prime}, m^{\prime} := K_{in}, K_{ex},m$ 
  \item $h^{\prime}_{older}(p,t) := \begin{cases}h_{older}(p,t)+d & iff \quad m \geq  (p,t)^{-} \\0 & iff \quad m \not\geq  (p,t)^{-} \end{cases}$
  \item $h_{after}^{\prime} := h_{after}^{\prime} + d$ 
\end{enumerate}

Note that the $h_{older}(p,t)$ delay marking does not begin to increment before the input place is populated with a token. For an older marking for input arc $(p,t)$ to release its on-chain transition $t$, it must increment beyond an earliest firing time $I_{older}(p, t)$, which in turn is determined by the presence of an older time-lock in the associated output satisfaction conditions given by $t.\pi_i$, where $\pi_i$ is the witness permutation associated with the on-chain transition $t$ (definition \ref{def: Trace-Net}). In the absence of an older time-lock, $I_{older}(p,t)$ is set to $0$. The $h_{after}(p,t)$ input arc time marking simply holds the value of the chain height at the time of evaluation and is initiated to the initial blockheight $b_0$. It releases the on-chain transition at the earliest firing time which is determined by the after time-lock of the output its input arc is connected to. In absence of an after time-lock, $I_{after}(p,t)$ is set to $0$. 

\begin{definition}{\rm\textbf{(Valid On-chain Transition)}} \label{def: Valid on-chain transaction}
An on-chain transition t is valid if
\end{definition}
\begin{enumerate}
  \item $m \geq  t^{-}$
  \item $h_{older}(p,t) \geq I_{older}(p,t))$ for all $p \in P.(p,t) \in F$ 
  \item $h_{after}(p,t) \geq I_{after}(p,t)$ for all $p \in P.(p,t) \in F$ 
\end{enumerate}

\begin{definition}{\rm\textbf{(Fire-able On-chain Transition)}} \label{def: Fire-able On-chain Transition}
A transition t in Trace-Net state Z is ready to fire by an actor if
\end{definition}
\begin{enumerate}
  \item $t$ is a valid on-chain transaction
  \item $t_{b}$ for $tx(t)$ is not fireable (definition: \ref{def: Transaction Broadcast Firing})
  \item $\exists actor (K_{actor} \vdash tx(t))$
\end{enumerate}

An on-chain transition $t$ can fire if its pre-places are populated, and all input arc time markings have released, and a single or both actors can deduce valid witnesses for all inputs from knowledge. A on-chain transition firing is proceeded by a transaction broadcast (definition \ref{def: Transaction Broadcast Firing}), if the broadcast results in an update of an actor's knowledge.

\begin{definition}{(On-chain transition firing)} \label{def: On-chain transition firing}
The firing of a fire-able on-chain transition of t in Trace-Net $\mathcal{Z}$ changes the state z into $z^{\prime} = (K_{in}, K_{ex}, h_{older}, h_{after}, m^{\prime})$ where
\end{definition}
\begin{enumerate}
  \item $m : = m + \Delta\:t$
\end{enumerate}
We denote $\Delta\:t$ as change in the place markings of the Trace-Net as a result of the on-chain transition firing.

\begin{definition}{\rm\textbf{(Chain roll-back transition firing)}} \label{def: Chain roll-back transition firing}
A roll-back of a blockchain by $n$ blocks by the adversarial external actor at state $z$ is always fireable. It reverts the transactions in the on-chain chronology which were confirmed within $n$ blocks of the current height $b$ and changes the contract state $z$ to $z^{\prime} = (K_{in}, K_{ex}, h_{older}^{\prime}, h_{after}^{\prime}, m^{\prime}, q^{\prime})$ where
\end{definition}
\begin{enumerate}
  \item $q^{\prime}$ has reverted all transactions in $q$ which were confirmed in the rolled-back blocks. 
  \item $m^{\prime} - \sum^{rev}\Delta t$, where $rev$ is the set of reverted transactions and $\Delta t$ the changes in place marking they effected.
  \item $h_{after}^{\prime}:=\begin{cases}h_{after} - n & if\; h_{after} - n > 0 \\0 & if \; h_{after} - n \leq 0\end{cases}$
  \item $h_{older}(p,t)^{\prime}:=\begin{cases}h_{older}(p,t) - n & if\; h_{older}(p,t) - n > 0 \\0 & if \; h_{after} - n \leq 0\end{cases}$
\end{enumerate}

Note that the knowledge of the actors is not rolled back during a roll-back transition firing, as the previously observed and derived knowledge of an actor cannot be deleted by the adversarial external actor. Once a roll-back is completed, it is possible to model alternative chain branches created by the 51\% attacker, who can generate any valid contract transaction order in the new branch (section \ref{section: Automated Trace-Net Analysis}).

%-------------------------------------------------------------------------------
\subsection{Trace-Net Generation}
%-------------------------------------------------------------------------------

Algorithm 1 summarizes the generation of a Trace-Net from a set of Bitcoin contract transaction templates in $K_{int,ext}$ after the completion of the implemented contract setup phase.

\begin{algorithm}[!h]
\caption{Trace-Net Generation} \label{Trace-Net Generation}
\begin{algorithmic}[1]
\STATE $P, T, F:= \emptyset , \emptyset, \emptyset$
\STATE $I_{older}, I_{after}, m_0 := \{\},\{\},\{\}$
\STATE $b_0 := current\;blockheight$
\STATE
\FORALL{$tx^{\prime} \in K_{in,ex}$}
\STATE
    \STATE \textit{/* Generate places for all outputs and prevouts */}
    \FORALL{$idx, in \in tx^{\prime}.\vec{ins}$}
        \STATE $\vec{p_{in}}[idx] := place(in.prevout)$
        \STATE Add $\vec{p_{in}}[idx]$ to $P$ if $\vec{p_{in}}[idx] \notin P$
        \STATE $m_0(\vec{p_{in}}[idx]) = 1$ if unspent on-chain
    \ENDFOR
    \FORALL{$idx, out \in tx^{\prime}.\vec{outs}$}
        \STATE $\vec{p_{out}}[idx] := place(out)$
        \STATE Add $\vec{p_{out}}[idx]$ to $P$ if $\vec{p_{out}}[idx] \notin P$
    \ENDFOR
    \STATE
    \STATE \textit{/* Generate permutations of $\pi(tx)$ over inputs */}
    \STATE $\langle\bar{\Gamma}\rangle := \bigcup_{in\in tx^{\prime}.ins}\{sat(in)\}$
    \STATE $\pi(tx) := product(\langle\bar{\Gamma}\rangle)$
    \STATE
    \STATE \textit{/* Per permutation: Construct transition, arcs */}
    \FORALL{$\pi_i \in \pi(tx)$}
        \STATE Add new $t$ to $T$
        \FORALL{$idx, p_{in} \in \vec{p_{in}}$}
            \STATE Add $(p_{in},t)$ to F
            \STATE $I_{older}(p_{in},t) := n_{older},\;s.t.\;n_{older} \models \pi_i$
            \STATE $I_{after}(p_{in},t) := n_{after},\;s.t.\;n_{after} \models \pi_i$
        \ENDFOR
        \FORALL{$p_{out} \in \vec{p_{out}}$}
            \STATE $arc_{out} : = arc(t,p_{out})$
            \STATE Add $arc_{out}$ to F
        \ENDFOR
    \ENDFOR
\STATE
\ENDFOR
\end{algorithmic}
\end{algorithm}

To generate the Trace-Net $\mathcal{Z}$ in state $z_0$, all transactions in the contract are iterated through. Places are generated for each in- and output in the contract (Lines 7-16). Subsequently, a transaction is represented by one or more on-chain transition instances, each representing a unique symbolic witness permutation: The \emph{product} returns a cartesian product over all sets of alternative symbolic witnesses $\bar{\Gamma}_i$ for each \emph{ith} input (Lines 19, 20) and returns a set of symoblic witness permutations $\pi(tx) = \{ \pi_0, \pi_1, ...\}$. A separate on-chain transition is instantiated for each with the corresponding input arcs adopting \emph{earliest firing time} values from the respective symbolic witness constraints (Lines 27, 28).

%-------------------------------------------------------------------------------
\section{Automated Trace-Net Analysis} \label{section: Automated Trace-Net Analysis}
%-------------------------------------------------------------------------------

%---------------------------x
\begin{figure*}[ht]
\begin{center}
\includegraphics[width=\textwidth]{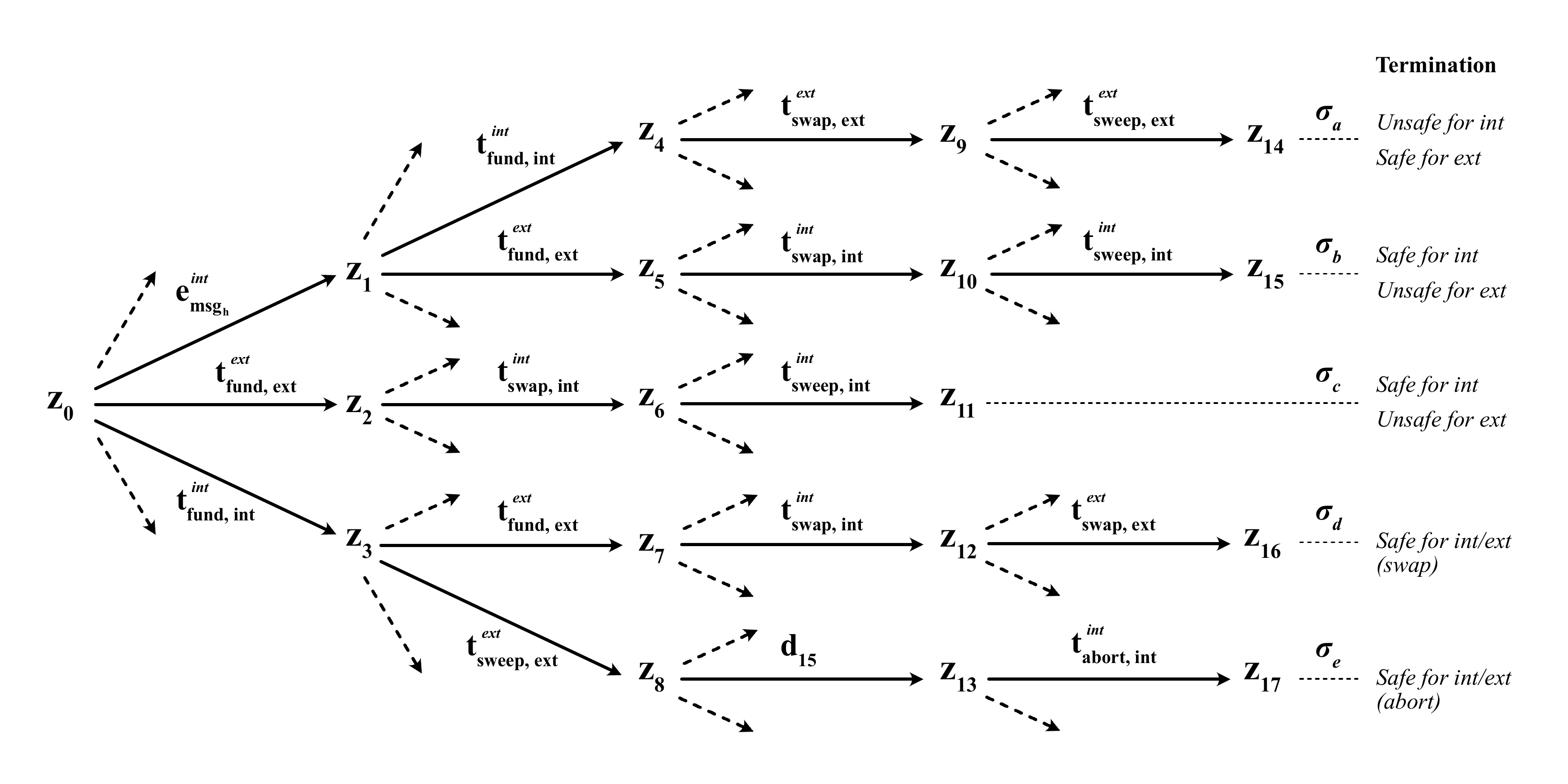}
\end{center}
\caption{A partial reachability graph for our atomic-swap contract example is shown above which has been generated with transaction confirmation delays and reorganization depths set to zero for simplicity. The contract is being evaluated for its initial state $z_{0}$, before any funding transaction has been executed. Each contract execution trace shown is denoted with the outcome of its execution.} 
\label{fig:reachabilitygraph}
\end{figure*}
%% %---------------------------

Let $\mathcal{Z}$ be our Trace-Net and $z_{0} = \{ K^{in}_0, K^{ex}_0, h_{older,0}=0, h_{after,0}=b_0, m_0, q_0 = \emptyset \}$ its initial state. We are interested in generating the transition system or \emph{reachability graph} $RG$ which expresses the total state space reachable by all possible contract executions. At each state $z$, the fireability of messages, transaction broadcast, transaction confirmation and blockchain roll-backs is evaluated, which in turn implies the strategies of each actor in state $z$. We parameterize the generation of $RG$ with two additional parameters: Confirmation delays and maximum blockchain reorganization depth.

\paragraph{Confirmation delays} The confirmation delay between on-chain transaction broadcast and confirmation in a block is denoted $c_{int}$ and $c_{ext}$ respectively, where $c_{int}, c_{ext} \in \mathbb{Z}$. Non-zero confirmation delays imply that an on-chain transaction fired by \emph{actor} will always be preceded by a delay transition of duration $c_{int/ext}$.

\paragraph{Blockchain reorganizations} \label{paragraph: blockchain reorgs}We explicitly model the ability of the adversary to reorganize the blockchain up to a depth of $r_{max} \in \mathbb{Z}$ when generating $RG$. At each Trace-Net state, alternate blockchain reorganizations of all depths up to $r_{max}$ are generated. An adversarial reorganization can be described as a $r$-depth \emph{roll-back} (definition \ref{def: Chain roll-back transition firing}) transition from state $z$ to state $z_{br}$, followed by the generation of all possible contract states reachable within $r+1$ blocks starting from $z_{br}$ fireable by the external actor (Alg. \ref{algo: RG Generation}, lines 26-33). Although not explored in alg. \ref{algo: RG Generation}, the chain heights from resulting reorganizations can include heights which are larger than 1. On-chain contract transitions fired by the external actor in the reorganized branch do not need to be prepended with a confirmation delay, as the external actor can directly append any valid transaction to any block in the reorganized chain, without having to broadcast to the public network. A blockchain reorganization from $z$ to the reorganized state $z'$ is a single externally fireable transition in the reachability graph: $z \xrightarrow{reorg} z^{\prime}$.

The resulting reachability graph $RG(\mathcal{Z}, z_{init}, c_{int,ext},r_{max}) = (W, E)$ is a directed graph with the set of vertices $W$, and edges $E$. It is necessarily finite, given the bounded knowledge expansion after each state transition and the finite number of contract transactions. Figure \ref{fig:reachabilitygraph} illustrates the partial reachability graph of an atomic swap contract initiated by the external actor. Algorithm \ref{algo: RG Generation} generates the contract state space reachable from an initial state $z_{0}$ by recursively computing all child states directly reachable via transitions fireable in each state. The time delay transition is only fired if its firing results in an expiration of time-locks, thereby releasing a contract transaction. The time-transition delay is chosen to be minimal, in order to reflect the transaction execution order enforced by contract time-locks. 

\begin{algorithm}[!h]
  \caption{Reachability Graph $RG(\mathcal{Z}, z_{init}, c_{int,ext}, r_{max})$} \label{algo: RG Generation}
  \begin{algorithmic}[1]
  \STATE $W, E := \left\{z_{init}\right\}, \emptyset$
  \FORALL{$z \in W$}
  \STATE
     \STATE \textit{/* States reachable by on-chain transitions */}
     \FORALL{$t$ fire-able in $z$}
        \STATE Compute $z^{\prime}$, $z^{\prime\prime}$, $z^{\prime\prime\prime}$ s.t. $z \overset{tb(t)}{\rightarrow} z^{\prime} \xrightarrow{d(c_{int/ext})} z^{\prime\prime} \overset{t}{\rightarrow}z^{\prime\prime\prime}$
        \STATE Add $z^{\prime}$, $z^{\prime\prime}$, $z^{\prime\prime}$ and  to $W$
        \STATE Add $\left(z,(tb),z^{\prime}\right)$, $\left(z^{\prime},(d),z^{\prime\prime}\right), \left(z^{\prime\prime},(t),z^{\prime\prime\prime}\right)$ to $E$
    \ENDFOR
    \STATE
    \STATE \textit{/* States reachable by message transitions */}
     \FORALL{$e$ fire-able in $z$}
        \STATE Compute $z^{\prime}$ such that $z \overset{e}{\rightarrow} z^{\prime}$
        \STATE Add $z^{\prime}$ and $\left(z,(e),z^{\prime}\right)$ to $W, E$
    \ENDFOR
    \STATE
    \STATE \textit{/* Timelock expiration(s) reachable with delays */}
    \FORALL{$t$ fire-able in $z$ after min. delay $d$}
      \STATE Compute $z^{\prime}$ such that $z \overset{d}{\rightarrow} z^{\prime}$
      \STATE Add $z^{\prime}$ and $\left(z,(d),z^{\prime}\right)$ to $W, E$
    \ENDFOR
    \STATE
    \STATE \textit{/* States reachable with chain reorganization(s) */}
    \FORALL{$dep$ in $[1:r_{max}]$}
        \STATE $W_r : = \emptyset$
        \STATE Compute $z_{br}$ s.t. $z \xrightarrow{r(dep)} z_{br}$
        \STATE Add $z_{br}$ to $W_r$
        \FORALL{$z_r$ in $W_r$}
            \FORALL{fireable $t^{ext}$/$d$ in $z_r$ s.t. $h(z_r^{\prime}) \leq h(z)+1$}
                \STATE Compute $z_r^{\prime}$ such that $z_r \xrightarrow{d} \xrightarrow{tb^{ext}} \xrightarrow{t^{ext}} z_r^{\prime}$
                \STATE Add $z_r^{\prime}$ to $W_r$, Del $z_r$ from $W_r$
            \ENDFOR
        \ENDFOR
        \STATE $W := W \cup W_r$
        \FORALL{$z_r$ in $W_r$}
            \STATE add $(z,reorg,z_r)$ to $E$
        \ENDFOR
    \ENDFOR
    \STATE
\ENDFOR
\end{algorithmic}
\end{algorithm}

\paragraph{Feasible termination trace} A feasible termination trace $\sigma$ will traverse the reachability graph $RG(\mathcal{Z}, z_0, c_{int,ext}, r_{max})$ from $z_{0}$ until a terminal state $z_{n}$ is reached, where no further transitions can be fired.
\[
  \sigma = z_{0} \overset{\theta_{0}}{\rightarrow} z_{1} ... z_{n-1}\overset{\theta_{n-1}}{\rightarrow}z_{n}
\]
The finite reachability graph allows an exhaustive analysis of all feasible contract traces. The types of transitions and firing actors can be analyzed to infer temporal contract safety properties of interest.

%-------------------------------------------------------------------------------
\section{Contract Safety Properties} \label{Contract Safety Properties}
%-------------------------------------------------------------------------------
%---------------------------

We can now formalize the notion of trustless contract execution, which implies that the verifying actor can always safely terminate the contract execution despite any adversarial strategies pursued by the external actor.

%-------------------------------------------------------------------------------
\subsection{Trustless Execution Property} \label{Trustless Execution Property}
%-------------------------------------------------------------------------------
%---------------------------

% \begin{figure}
% \begin{center}
% \includegraphics[width=0.47\textwidth]{trace_safety.jpg}
% \end{center}
% \caption{ A safe contract exists at state $z_{i}$ if an internal actor can unilaterally execute a path which leads to a terminally safe state ($z_{s}, z_{s}^{\prime}, z_{s}^{\prime\prime}$). This path must also be safe against competing transitions fired ($t_{1b}^{ext}, t_{2c}^{ext}$) by the external actor.}
% \label{fig:contractsafety}
% \end{figure}
% %---------------------------

\paragraph{(i) Safe Terminal States} The safety of terminal states in a reachability graph $RG(\mathcal{Z}, z_{init},..)$ are determined by a user-supplied safety policy function $\textsf{policy}(K_{int,ext}, m)$, and therefore depends on the actor knowledge state and output markings at the terminal state in question. Actor knowledge states can imply information leaked to the counter-party, and place markings infer the balance owned by the internal actor. For our atomic swap example, we could suggest a policy function which ensures that in a safe terminal state, the internal actor must obtain a balance equal to the amount used to fund the contract, thereby ensuring that only terminal states of the abort and success paths are considered safe. 

\paragraph{(ii) (Un)cooperative Termination Trace} An execution trace $\sigma^{int}$ in $RG$ beginning at a state $z_{init}$ which leads to a safe terminal state within finite transitions fired by a single actor is an uncooperative termination trace. The presence of internally fireable, uncooperative traces from contract state $z$ implies that the verifying actor can reach a safe terminal state if the external actor ceases to participate in the contract protocol. A collaborative termination trace is a contract execution trace which includes transitions fired by both actors.  

\paragraph{(iii) Safe Non-terminal States} 
\begin{enumerate}
  \item A safe non-terminal state $z_{s}$ in $RG$ features at least one uncooperative termination trace $\sigma^{int}$ fireable by the verifying actor.
  \item If the external actor can fire any on-chain transition $t^{ext}$ in $z$ along $\sigma^{int}$, then the resulting state $z^{\prime}$ computed from $z \overset{t^{ext}}{\rightarrow} z^{\prime}$ must also be a safe state. 
\end{enumerate}

\paragraph{(iv) Safe Termination Trace} If all intermediary states along a (un)cooperative termination trace $\sigma^{int/ext}$ are safe for the verifying internal actor, then the trace is considered safe, as the verifying actor is guaranteed to have an execution path to safe termination, despite any adversarial strategies pursued by the actor. 

% \paragraph{(iv) Trustless Execution} 

% Throughout the life-cycle of a contract, actors may negotiate to collaboratively transition to different contract states. The internal actor must verify that each new contract state is a safe non-terminal state $z_{i}$ with a safe contract termination trace $\sigma_{s}$.

\begin{definition}{\rm\textbf{(Trustless Execution Property)}} \label{def: Trustless Execution Property}
The contract execution from a state $z_{init}$ exhibits the trustless execution property if there exists at least one safe cooperative trace beginning at $z_{init}$.
\end{definition}

%-------------------------------------------------------------------------------
\subsection{Analysis of Contract Updates} \label{Safe Contract Updates}
%-------------------------------------------------------------------------------

A contract update can occur whilst a contract is not yet terminated. This implies the execution of an additional setup phase and new transaction templates being added the current actor knowledge $K_{int}, K_{ext}, RG \xrightarrow{update} K_{int}^{\prime}, K_{ext}^{\prime}, RG^{\prime}$, from which an updated Trace-Net and reachability graph $RG^{\prime}$ are generated. If $RG^{\prime}$ features additional, safely reachable terminal states the contract update is safe, as it increases the contract-state space with additional safe termination traces.

%-------------------------------------------------------------------------------
\subsection{Contract State Stability} \label{Contract State Stability}
%-------------------------------------------------------------------------------

A contract state $z$ is stable if
\begin{equation}
RG(\mathcal{Z}, z, c_{int,ext}, r_{max}) = RG(\mathcal{Z}, z^{\prime}, c_{int,ext}, r_{max})
\end{equation}

where $z \overset{\infty}{\rightarrow} z^{\prime}$ and $\infty$ is a delay of infinite duration. This implies that the execution of the contract can be deferred indefinitely. Both the analysis of contract updates and verification of the state stability property can be useful in the analysis of off-chain protocols, which defer transaction broadcasts so that contract updates can be performed off-chain.

%-------------------------------------------------------------------------------
\section{Conclusion}
%-------------------------------------------------------------------------------
We have introduced a novel method to automate the verification of multi-party Bitcoin contract protocols at the raw contract transactions level. The main advantages of contract verification at the implementation level is the ability to verify a contract at run-time and to accurately model the underlying blockchain execution environment, which may include confirmation delays and possible chain reorganizations. Furthermore, Trace-Net can be useful as an effective monitoring framework for contract protocol implementations.

The main components of our framework consist of a stateful actor knowledge model and an extended Petri Net model which captures the semantics of on-chain contract output availability and time-locks. Together, these two components determine which strategies are available to contracting participants in all contract protocol states. Furthermore, we have demonstrated the extension of Trace-Net actor knowledge model with \emph{adaptor signatures} in section \ref{Cryptographic Extensions}, thereby extending Trace-Net verification to contracts which can be executed privately.

The instantiation of Trace-Net was illustrated with \emph{atomic swaps} and we emphasize that Trace-Net semantics are sufficiently expressive for more complex protocols such as payment channel networks \cite{poon16bitcoin} or generalized state-channels \cite{aumayr2020generalized}. Trace-Net lends itself to further investigations, including methods to efficiently represent the unfolded state-space in which the verification is performed, as the state-space explosion problem remains untreated. Proposals \cite{chakravarty2020extended} \cite{bartoletti2020bitcoin} have been made to extend Bitcoin script, providing Turing-machine expressiveness, which would be a valuable future area of research for Trace-Net. Furthermore, domain-specific contract specification languages can be investigated which compile to Trace-Net models as an intermediate representation, whilst providing inherent contract safety and liveness guarantees. We intend to develop research tooling to explore the verification of Trace-Net models with time-proven model-checking frameworks.

\section*{Acknowledgments}
Thanks to the reviewers of the early versions of this paper. A special acknowledgement to Josh Harvey for early discussions on formal Petri Net specifications of automata-based programs which inspired the idea of Trace-Net for Bitcoin contracts. Bitcoin contracting protocol experts Antoine Riard, Dmitry Petukhov, Nadav Cohen and Thibaut le Guilly all provided valuable perspectives to refine Trace-Net semantics. Finally, the author wishes to express his sincere gratitude to Professor Alberto Lluch Lafuente and Professor Andrea Vandin for graciously sharing their expertise in formal methods and model-checking methodology, without which this work would not have been possible.
\bibliographystyle{unsrt}
\bibliography{paper}

%%%%%%%%%%%%%%%%%%%%%%%%%%%%%%%%%%%%%%%%%%%%%%%%%%%%%%%%%%%%%%%%%%%%%%%%%%%%%%%%
\end{document}